\documentclass[12pt]{article}
\usepackage {graphicx}
\newcommand{\dfrac}{\displaystyle\frac}
\newcommand{\dsum}{\displaystyle\sum}
\newcommand{\nn}{\nonumber}
\newcommand{\lb}{\linebreak}
\newcommand{\q}{\quad}
\newcommand{\qq}{\qquad}

\title{ON THE THEORY OF INTERACTING FIELDS\\
IN FOLDY-WOUTHUYSEN REPRESENTATION\footnote{This paper was written
by the author in 1988 and executed as a report of RFNC-VNIIEF. It
is the author's view that the results of the efforts continue to
be
topical, so they are offered to the reader in the form of paper.}}
\author{V. P. Neznamov}
\begin{document}
 \maketitle

\begin{abstract}
The paper considers quantum electrodynamics (QED) and weak
interaction of elementary particles in the lower orders of the
perturbation theory using nonlocal Hamiltonian $H_{FW}$ in the
Foldy-Wouthuysen ($FW$) representation. Feynman rules in the $FW$
representation are specified, specific QED processes are
calculated. Cross sections of Coulomb scattering of electrons,
M\"oller scattering, Compton effect, electron self-energy, vacuum
polarization, anomalous magnetic moment of electron, Lamb shift of
atomic energy levels are calculated. The possibility of the
scattering matrix expansion in powers of the coupling constant, in
which matrix elements $S_{fi}$ contain no terms with fermion
propagators, is demonstrated for external fermion lines
corresponding to real particles (antiparticles).

It is shown that a method to include the interaction of real
particles with antiparticles in the $FW$ representation is to
introduce negative mass particles and antiparticles to the theory.
The theory is degenerate with respect to the particle
(antiparticle) mass sign, however the masses of the particle and
antiparticle interacting with each other should be of opposite
sign.

QED in the $FW$ representation is invariant under $C$, $P$, $T$
inversions. The weak interaction breaks the $C$ and $P$
invariance, but preserves the combined $CP$ parity. In the theory
there is a possibility to relate the break of $CP$ invariance to
total or partial removal of the degeneracy in particle
(antiparticle) mass sign.
\end{abstract}
\newpage


 \section*{INTRODUCTION}

Historically, the first equation describing the $\dfrac12$-spin
particle with electromagnetic field has been non-relativistic
Pauli equation
\begin{equation}
\label{eq1} p_{0} \varphi \left( {x} \right) = \left[
{\dfrac{{\left( {\vec {p} - e\vec {A}\left( {x} \right)}
\right)^{2}}}{{2m}} - \dfrac{{e\vec {\sigma} \vec {B}}}{{2m}} +
eA_{0} \left( {x} \right)} \right]  \varphi \left( {x} \right)
\end{equation}

In (\ref{eq1}) and hereinafter the system of units $\hbar = c = 1$
is used; $x$, $p$, $A$ are 4-vectors;   $xy =\lb= x^{\mu} y_{\mu}
= x^{0}y^{0} - x^{k}y^{k}$, $\mu = 0,1,2,3$,  $k = 1,2,3$ as
usual; $p^{\mu} = i\dfrac{{\partial} }{{\partial x^{\mu} }}$;
$\vec {B} = rot\vec {A}$; $\sigma ^{k}$ are Pauli matrices;
$\varphi \left( {x} \right)$ is a two-component wave function.

Then Dirac derived his famous relativistic equation describing the
$\dfrac{1}{2}$-spin particle motion. Given the interaction with
magnetic field, the Dirac equation takes the form:
\begin{equation}
\label{eq2} p_{0} \psi _{D} \left( {x} \right) = \left[ {\vec
{\alpha} \left( {\vec {p} - e\vec {A}\left( {x} \right)} \right) +
\beta m + eA_{0} \left( {x} \right)} \right]  \psi _{D} \left( {x}
\right),
\end{equation}
 $\psi _{D} \left( {x} \right)$ ---  is the four-component wave function,
$\alpha ^{i} = \left( {\begin{array}{cc}
 0   &\sigma ^{i} \\
 \sigma ^{i} &0
 \end{array}}\right)$,
  $\beta = \left( {\begin{array}{cc}
 I &0 \\
 0  &-I
 \end{array}} \right)$ are the Dirac matrices. The Dirac
equation, unlike equation (\ref{eq1}), is linear in components of
impulse $p^{\mu} $ and is readily representable in the explicitly
covariant form. The Pauli equation is a nonrelativistic limit of
the Dirac equation for the upper components of the wave function
$\psi _{D} \left( {x} \right)$.

In principle, the relativistic generalization of the Pauli
equation could be performed in a different way with using the
relativistic relation between energy and particle impulse as a
basis when there are no external fields. This was actually done by
Foldy and Wouthuysen in their classic paper \cite{1}. The
Foldy-Wouthuysen equation for free motion is
\begin{equation}
\label{eq3} p_{0}  \psi _{FW} \left( {x} \right) = \left( {H_{0}}
\right)_{FW} \Psi _{FW} \left( {x} \right) = \beta E\Psi _{FW}
\left( {x} \right)
\end{equation}

In (\ref{eq3})  $E = \sqrt {\vec {p}^{\,2} + m^{2}} $.

Solutions to (\ref{eq3}) are plane waves of positive and negative
energies
\begin{equation}
\label{eq4} \psi _{FW}^{\left( { +}  \right)} \left( {x,s} \right)
= \dfrac{{1}}{{\left( {2\pi}  \right)^{3/2}}}U_{s} e^{ - ipx},\q
\psi _{FW}^{\left( { -}  \right)} \left( {x,s} \right) =
\dfrac{{1}}{{\left( {2\pi}  \right)^{3/2}}}V_{s} e^{ipx},\q  p_{0}
= \left( {\vec {p}^{\,2} + m^{2}} \right)^{1/2}
\end{equation}

In (\ref{eq4})
 $U_{s} = \left( \begin{array}{c}
 \chi_{s}\\
 0
 \end{array}\right)$, $V_s=\left( \begin{array}{c}
 0\\[-2pt]
 \chi_{s}
 \end{array} \right)$, $\chi_s$ are two-component normalized Pauli
spin functions.

The following orthonormalization-completeness relationships are
valid for $U_{s} $ and $V_{s} $:
\begin{equation}
\label{eq5}
\begin{array}{l}
 U_{s}^{ +}  U_{s^{'}} = V_{s}^{ +}  V_{s^{'}} = \delta _{ss^{'}} ;\q U_{s}^{
+}  V_{s^{'}} = V_{s}^{ +}  U_{s^{'}} = 0; \\
 \dsum\limits_{s} {} \left( {U_{s}}  \right)_{\gamma}  \left( {U_{s}^{ +} }
\right)_{\delta}  = \dfrac{{1}}{{2}}\left( {I+\beta}
\right)_{\gamma \delta}  ;\quad \dsum\limits_{s} {} \left( {V_{s}}
\right)_{\gamma}  \left( {V_{s}^{ +} }  \right)_{\delta}  =
\dfrac{{1}}{{2}}\left( {I-\beta} \right)_{\gamma \delta}.  \\
 \end{array}
\end{equation}

In (\ref{eq4}), (\ref{eq5}) $\gamma ,\delta $ refer to the spinor
subscripts, $s$ to the spin ones. In (\ref{eq5}) and below the
summation symbol and the subscripts themselves are omitted in the
summation over the spinor subscripts.

Hamiltonian $\left( {H_{0}}  \right)_{FW} $ is related with free
Dirac Hamiltonian $\left( {H_{0}}  \right)_{D} $ by the unitary
transformation. In equation (\ref{eq3}) definite asymmetry of
space coordinates and time is seen, although it is
Lorentz-invariant by itself.

In the general case of the interaction with external
electromagnetic field $A^{\mu} \left( {x} \right)$ there is no
exact unitary transformation that would transform Dirac equation
(\ref{eq2}) to Foldy-Wouthuysen $\left( {FW} \right)$
representation. In this case Foldy and Wouthuysen found
Hamiltonian $\left( {H_{FW}}  \right)$ in the form of series in
terms of powers of $1/m$ \cite{1}.

Blount \cite{2} found Hamiltonian $\left( {H_{FW}}  \right)$ in
the form of series in terms of powers of the smallness of fields
and their time and space derivatives. Case \cite{3} obtained an
accurate transformation in the presence of time-independent
external magnetic field $\vec {B} = rot\vec {A}$. In this case the
Dirac equation is transformed to equation
\begin{equation}
\label{eq6} p_{0} \psi _{FW} \left( {x} \right) = H_{FW} \psi
_{FW} \left( {x} \right) = \beta \sqrt {\left( {\vec {p} - e\vec
{A}} \right)^{2} - e\vec {\sigma} \vec {B} + m^{2}} \cdot \psi
_{FW} \left( {x} \right).
\end{equation}

The author   obtains the relativistic Hamiltonian $\left( {H_{FW}}
\right)$ in the form of series in terms of powers of charge $e$ in
the general case of interaction with external field $A^{\mu}
\left( {x} \right)$ \cite{4}.

In all the cases considered, in the Foldy-Wouthuysen
representation the equations for wave functions are of
noncovariant form and their Hamiltonians are nonlocal. In the $FW$
representation there is no relation between the upper and lower
wave function components, that is, $H_{FW} $ is essentially a
two-component Hamiltonian.

With the advent of the relativistic Hamiltonian $\left( {H_{FW}}
\right)$ obtained in ref. \cite{4}, it became possible to consider
the $FW$ representation of quantum-field processes in the context
of the perturbation theory. Why is it interesting irrespective of
the non-covariance of the Dirac equation in the Foldy-Wouthuysen
representation, nonlocality, and relative complexity of the
expressions for Hamiltonian?

First, a number of paradoxes characteristic of the Dirac
representation are resolved in the $FW$ representation. In the
$FW$ representation, in the free case, the velocity operator is of
habitual form $\vec {V} = \dfrac{{\vec {p}}}{{E}}$, which is close
to the classic form (in the Dirac representation, $\vec {V} = \vec
{\alpha} $), there is no electron "jitter" (Zitterbewegung), the
spin operator remains unchanged in time \cite{1}.

Second, since in Hamiltonian $H_{FW} $ there is no relation of the
initial (final) states to positive energy and of final (initial)
states to negative energy, in the quantum field theory there will
be no interactions of real particle-antiparticle pairs. To include
these processes, additional terms must be introduced to the
Hamiltonian $H_{FW} $.

In this connection the consideration of the quantum field theory
in the $FW$ representation can lead to new physical consequences
and re-interpretation of the habitual terms in the Dirac
representation.

Section 1 of this paper is devoted to quantum electrodynamics in
the Foldy-Wouthuysen representation. Here Hamiltonian $H_{FW} $ is
given in the form of series in terms of powers of $e$, Feynman
rules in the$FW$ representation are specified, results of
calculations of specific QED processes are discussed. For the
case, where the external fermion lines correspond to real
particles (antiparticles), the possibility of the scattering
matrix expansion in powers of $e$, in which scattering matrix
elements \textit{S}$_{fi}$ contain no terms with electron-positron
propagators, is demonstrated.

The formulas for vertex operators are therewith close in their
structure to those of the "old", noncovariant perturbation theory
developed by Heitler \cite{5}.

Next, Section 1 demonstrates that a method to include the external
electron-positron pair interaction in the $FW$ representation is
to introduce the relationship between the solutions to the
$FW$-transformed Dirac equations with positive and negative
particle (antiparticle) mass. In so doing the positive-mass
particles (antiparticles) interact with one another and with
negative-mass antiparticles (particles). Conversely, negative-mass
particles (antiparticles) interact with one another and with
positive-mass antiparticles (particles). The theory is degenerate
with respect to the particle (antiparticle) mass sign, however the
masses of the particle and antiparticle interacting with each
other should be of opposite signs. It should be particularly
emphasized that in the context under discussion the particle mass
sign is an internal quantum number not related by the author in
this paper to the gravitational interaction sign. Of course, the
negative-mass particle energy is therewith positive. Note that
previously the conclusion of the opposite signs in the particle
and antiparticle masses was made by Recami and Ziino \cite{6} when
analyzing the special relativity theory and conditions of
reversibility "particle $ \leftrightarrow $ antiparticle". The
suggested form of QED in the $FW$ representation is invariant
under $P,C,T$ transformations.

Section 2 is devoted to the theory of weak interaction in the
$FW$representation. The weak interaction is considered in the
lower order of the perturbation theory, in the $V$-$A$ form of the
four-fermion interaction. The weak current in the $FW$
representation is determined explicitly. Like in the Dirac
representation, weak interactions break the $P$ and $C$
invariance, but preserve the combined $CP$ parity. In the theory
there is a possibility to relate the $CP$ invariance break to the
partial removal of the degeneracy in particle (antiparticle) mass
sign.

\section{QUANTUM ELECTRODYNAMICS IN THE\\
FOLDY-WOUTHUYSEN REPRESENTATION\\ \cite{4}, \cite{7}, \cite{8}}
\subsection {Interaction Hamiltonian in the $FW$ representation}

In notation of ref. \cite{4} Dirac equation for quantized
electron-positron field in the $FW$ re\-presentation is written as
\begin{equation}\label{eq7}
\begin {array}{l}
 p_{0} \psi _{FW} \left( {x} \right) = H_{FW} \psi
_{FW} \left( {x} \right) = \left( {\beta E + K_{1} + K_{2} + K_{3}
+ \ldots} \right)\psi _{FW} \left( {x} \right),\\ K_{1} \sim e,\q
K_{2} \sim e^{2},\q K_{3} \sim e^{3}.
\end{array}
\end{equation}

In the impulse representation, with electromagnetic field
expansion in plane waves
\[
A_{\mu}  \left( {\vec {x},t} \right) = \sum\limits_{\nu = \pm 1} {} \int
{A_{\mu k}^{\left( {\nu}  \right)} e^{i\nu k_{o} t}e^{ - i\vec {k}\vec
{x}}d\vec {k}}
\]
\noindent matrix elements of interaction Hamiltonian terms $K_{1}
,K_{2} $ can be written as
\begin{eqnarray}
\label{eq8} && \langle \vec {p}^{\,'}\left| {K_{1}}  \right|{\kern
1pt} \vec {p}'' \rangle = \sum\limits_{\nu = \pm 1} {}
\dfrac{{\beta E^{''} - \beta E^{'} - \nu k_{0} }}{{R^{'} +
R^{''}}}\left( { - \dfrac{{1}}{{\beta E^{'} + \beta E^{''} - \nu
k_{0}} } \langle \vec {p}^{\,'}\left| {RLN} \right|\vec {p}^{\,''}
\rangle -}  \right. \nn\\
 &&\left. - \dfrac{{1}}{{\beta E^{'} + \beta E^{''} + \nu k_{0}} } \langle \vec
{p}^{\,'}\left| {NLR} \right|\vec {p}^{\,''} \rangle
 \right)+ \langle \vec {p}^{\,'}
 \left| C \right|\vec {p}^{\,''} \rangle;\\
\label{eq9} && \langle \vec {p}^{\,'}\left| {K_{2}}  \right|\vec
{p}^{\,''} \rangle = \sum\limits_{\nu ,\nu ^{'} = \pm 1} {} \int
{d\vec {p}^{\,'''}\left\{ {\dfrac{{\beta E^{''} - \beta E^{'} -
\nu k_{0} - \nu ^{'}k_{0}^{'}} }{{R^{'} + R^{''}}} \times}
\right.} \nn\\ && \times\left[ \dfrac{1}{\left( {\beta E^{'} +
\beta E^{'''} - \nu k_{0}} \right)\left( {\beta E^{''} - \beta
E^{'''} - \nu ^{'}k_{0}^{'}} \right)}\right.  \langle \vec
{p}^{\,'}\left| R\left( {C + \xi} \right) \right|\vec {p}^{\,'''}
\rangle \langle \vec {p}^{\,'''}\left| K_{1} \right|\vec
{p}^{\,''} \rangle + \nn\\ && + \dfrac{1}{\left( {\beta E^{''} +
\beta E^{'''} + \nu ^{'}k_{0}^{'}} \right)\left( {\beta E^{'''} -
\beta E^{'} - \nu ^{}k_{0}^{}}  \right)} \langle \vec
{p}^{\,'}\left| K_{1} \right|\vec {p}^{\,'''} \rangle \langle \vec
{p}^{\,'''}\left| \left( {C + \xi ^{ +} } \right)\left. {R}
\right| \vec {p}^{\,''} \right. \rangle - \nn\\ &&
  - \dfrac{1}{\left( {\beta E^{''} + \beta E^{'} - \nu k_{0} - \nu
^{'}k_{0}^{'}}  \right)\left( {\beta E^{''} + \beta E^{'''} - \nu
^{'}k_{0}^{'}}  \right)} \langle \vec {p}^{\,'}\left| R\left( {N +
\varphi} \right) \right|\vec {p}^{\,'''} \rangle \langle \vec
{p}^{\,'''}\left| N \right| \vec {p}^{\,''} \rangle + \nn\\ &&  +
\dfrac{1}{\left( {\beta E^{''} + \beta E^{'} + \nu k_{0} + \nu
^{'}k_{0}^{'}} \right)\left( {\beta E^{'''} + \beta E^{'} + \nu
^{}k_{0}^{} } \right)} \langle \vec {p}^{\,'}\left| N \right|\vec
{p}^{\,'''} \rangle \langle \vec {p}^{\,'''}\left| \left( {N +
\varphi ^{ +} } \right)R \right| \vec {p}^{\,''} \rangle - \nn\\
&&
  - \dfrac{1}{\left( {\beta E^{''} + \beta E^{'} - \nu k_{0} - \nu
^{'}k_{0}^{'}}  \right)\left( {\beta E^{''} - \beta E^{'''} - \nu
^{'}k_{0}^{'}}  \right)} \langle \vec {p}^{\,'}\left| R\left( {C +
\xi} \right) \right|\vec {p}^{\,'''} \rangle \langle \vec
{p}^{\,'''}\left| C \right| \vec {p}^{\,''} \rangle - \nn\\ &&
\left. - \dfrac{1}{\left( {\beta E^{''} + \beta E^{'} + \nu k_{0}
+ \nu ^{'}k_{0}^{'}} \right)\left( {\beta E^{'''} - \beta E^{'} -
\nu ^{}k_{0}^{} } \right)} \langle \vec {p}^{\,'}\left| C
\right|\vec {p}^{\,'''} \rangle \langle \vec {p}^{\,'''}\left|
\left( {C + \xi ^{ +} } \right)R \right| \vec {p}^{\,''} \rangle
 \right] + \nn\\
 &&+\dfrac{1}{R^{'} + R^{''}}\left[ -
\dfrac{{\beta E^{''} - \beta E^{'} - \nu k_{0} - \nu
^{'}k_{0}^{'}} }{{\left( {\beta E^{'''} - \beta E^{'} - \nu k_{0}}
\right)\left( {\beta E^{''} - \beta E^{'''} - \nu ^{'}k_{0}^{'}}
\right)}} \right. \langle \vec {p}^{\,'}\left| RC \right|\vec
{p}^{\,'''} \rangle \langle \vec {p}^{\,'''}\left| K_{1} \right|
\vec {p}^{\,''} \rangle + \nn\\ && + \dfrac{\beta E^{''} - \beta
E^{'} - \nu k_{0} - \nu ^{'}k_{0}^{'} }{\left( {\beta E^{'''} -
\beta E^{'} - \nu k_{0}} \right)\left( {\beta E^{''} - \beta
E^{'''} - \nu ^{'}k_{0}^{'}} \right)} \langle \vec {p}^{\,'}\left|
K_{1} \right|\vec {p}^{\,'''} \rangle \langle \vec
{p}^{\,'''}\left| CR \right| \vec {p}^{\,''} \rangle + \nn\\ &&  +
\dfrac{1}{\beta E^{'''} - \beta E^{'} - \nu k_{0}}  \langle \vec
{p}^{\,'}\left| RK_{1} \right|\vec {p}^{\,'''} \rangle \langle
\vec {p}^{\,'''}\left| K_{1} \right| \vec {p}^{\,''} \rangle -
\nn\\ &&  - \dfrac{1}{\beta E^{''} - \beta E^{'''} - \nu
^{'}k_{0}^{'}} \langle \vec {p}^{\,'}\left| K_{1} \right|\vec
{p}^{\,'''} \rangle \langle \vec {p}^{\,'''}\left| K_{1} R \right|
\vec {p}^{\,''} \rangle + \nn\\ && + \dfrac{1}{\beta E^{''} -
\beta E^{'''} - \nu ^{'}k_{0}^{'}} \langle \vec {p}^{\,'}\left| RC
\right|\vec {p}^{\,'''} \rangle \langle \vec {p}^{\,'''}\left|C
\right| \vec {p}^{\,''} \rangle - \nn\\ &&  -   \dfrac{1}{\beta
E^{'''} - \beta E^{'} - \nu k_{0}} \langle \vec {p}^{\,'}\left| C
\right|\vec {p}^{\,'''} \rangle \langle \vec {p}^{\,'''}\left| CR
\right| \vec {p}^{\,''} \rangle + \nn\\ && + \dfrac{1}{\beta
E^{''} + \beta E^{'''} - \nu ^{'}k_{0}^{'}} \langle \vec
{p}^{\,'}\left| RN \right|\vec {p}^{\,'''} \rangle \langle \vec
{p}^{\,'''}\left| N \right| \vec {p}^{\,''} \rangle + \nn\\ &&
\left.\left. + \dfrac{1}{\beta E^{'''} + \beta E^{'} + \nu k_{0}}
\langle \vec {p}^{\,'}\left| N \right|\vec {p}^{\,'''} \rangle
\langle \vec {p}^{\,'''}\left| NR \right| \vec {p}^{\,''} \rangle
\right]\right\}.
 \end{eqnarray}
In (\ref{eq9})
 $\xi = - \dfrac{{1}}{{R}}\left( {A_{0} + \vec
{\alpha} \vec {A}\dfrac{{\beta \vec {\alpha} \vec {p}}}{{E + m}}}
\right)   R,\qq
 \varphi = \dfrac{{1}}{{R}}\left( {A_{0} \dfrac{{\beta \vec {\alpha} \vec
{p}}}{{E + m}} + \vec {\alpha} \vec {A}} \right)   R$,\\ \noindent
$\xi ^{ +} ,\varphi ^{ +} $ are operators Hermitean conjugate to
$\xi ,\varphi $.

In (\ref{eq8}) and (9) operators $C,\,N,\,L,\,R$ are
\[
\begin{array}{l}
 C = eR\left( {A_{0} - L  A_{0}  L} \right)R - eR\left( {L
\vec {\alpha} \vec {A} - \vec {\alpha} \vec {A}  L} \right)R,
\\
 N = eR\left( {LA_{0} - A_{0}  L} \right)R - eR\left( {\vec {\alpha
}\vec {A} - L\vec {\alpha} \vec {A}  L} \right)R, \\
 L = \dfrac{{\beta \vec {\alpha} \vec {p}}}{{E + m}}, \quad R = \sqrt {\dfrac{{E
+ m}}{{2E}}} .
 \end{array}
\]

The expressions for $C,\,R$ are even operators, that is the
operators not relating the upper and lower components of wave
functions $\Psi _{FW} \left( {x} \right)$; accordingly, the
expressions for $N,L$ are odd operators. The interaction
Hamiltonian terms $K_{1} ,K_{2}, \ldots$, as they must, are even
operators.

An algorithm for determination of the $K_{1} ,K_{2} $ and the
following terms of the expansion in powers of $e$ is given in ref.
\cite{4}. Appendix 1 of this paper briefly presents the algorithm
for determination of Hamiltonian terms $K_{1}$, $K_{2}$, $K_{3}$.

\subsection{Feynman rules in the $FW$ representation}

The Feynman propagator of the Dirac equation in the
Foldy-Wouthuysen representation is
\begin{eqnarray}
\label{eq10} && S_{FW} \left( {x - y} \right) =
\dfrac{{1}}{{\left( {2\pi} \right)^{4}}}\int {d^{4}p\dfrac{{e^{ -
i\rho \left( {x - y} \right)}}}{{\rho _{0} - \beta E}}} =
\dfrac{{1}}{{\left( {2\pi} \right)^{4}}}\int {d^{4}p  } e^{ -
i\rho \left( {x - y} \right)}  \dfrac{{p_{0} + \beta E}}{{p^{2} -
m^{2} + i\varepsilon} } = \nn\\ && = - i\theta \left( {x_{0} -
y_{0}}  \right)\int {d\vec {p}\sum\limits_{s} {\psi _{FW}^{\left(
{ +}  \right)} \left( {x,s} \right)  \left( {\psi _{FW}^{\left( {
+}  \right)} \left( {y,s} \right)} \right)}} ^{ +} + \nn\\ && +
i\theta \left( {y_{0} - x_{0}}  \right)\int {d\vec
{p}\sum\limits_{s} {\psi _{FW}^{\left( { -}  \right)} \left( {x,s}
\right)  \left( {\psi _{FW}^{\left( { -} \right)} \left( {y,s}
\right)} \right)}}  ^{ +}.
 \end{eqnarray}

Eq. (\ref{eq10}) implies the Feynman rule of pole bypass;
$
\theta \left( {x_{0}}  \right) = \left\{\begin{array}{ll} 1, &
x_{0}> 0;\\ 0, & x_{0}<0. \end{array}\right. $

The integral equation for $\psi _{FW} \left( {x} \right)$ is
\begin{equation}
\label{eq11} \psi _{FW} \left( {x} \right) = \psi _{0} \left( {x}
\right) + \int {d^{4}yS_{FW} \left( {x - y} \right) \left( {K_{1}
+ K_{2} + \ldots} \right)  \psi _{FW} \left( {y} \right)}.
\end{equation}

In (\ref{eq11}) $\psi _{0} \left( {x} \right)$ is the solution to
the Dirac equation in the $FW$representation in the absence of
electromagnetic field $\left( {A^{\mu}  = 0} \right).$

Expressions (\ref{eq10}), (\ref{eq11}) allow us to formulate the
Feynman rules for recording elements of scattering matrix $S_{fi}
$ and calculating QED processes. In contrast to the Dirac
representation, in the \textit{FW} representation there is an
infinite set of types of photon interaction vertices depending on
the perturbation theory order: the vertex of interaction with one
photon is correspondent with factor $\left( { - i(K_{1})_\mu }
\right)$, the vertex of interaction with two photons with factor
$\left( { - i(K_{2})_{\mu \nu} } \right)$ and so on. For
convenience the corresponding parts of the terms of interaction
Hamiltonian $K_{1} ,K_{2},\ldots$ without electromagnetic
potentials $A^{\mu} ,A^{\mu} A^{\nu} ,\ldots$ are denoted by
$(K_{1})_\mu ,(K_{2})_{\mu \nu} ,\ldots$.

Each external fermion line is correspondent with one of functions
(\ref{eq4}). As usually, positive energy solutions correspond to
particles, negative energy solutions to antiparticles. The other
Feynman rules remain the same as those in the spinor
electrodynamics in the Dirac representation.

\subsection{Calculations of QED processes in the
$FW$ representation}

The formulated Feynman rules have been used to consider some QED
processes in the first and second orders of the perturbation
theory. Cross sections of Coulomb scattering of electrons,
M\"oller scattering, Compton effect, electron self-energy, vacuum
polarization, anomalous magnetic moment of electron, Lamb shift of
atomic energy levels have been calculated. Below are given the
Feynman diagrams of the processes considered. The brief details of
the calculations can be found in Appendix 2.

The final results of calculations of the QED processes, whose
diagrams are presented in Figs. 1---5, agree with the relevant
data calculated in the Dirac representation. The radiation
corrections to the electron scattering in the external field (Fig.
6) in the mass and charge renormalization provide a proper value
of the anomalous magnetic electron and Lamb shift of energy
levels.

A feature of the theory is that in the terms of the interaction
Hamiltonian $K_{n} $ (except $K_{1} $) there is an even number
\textit{N} of odd operators relating the initial and final states
of positive energy to the intermediate states of negative energy
and vice versa. Thanks to this, for example, the diagram of Fig.~5
appears that relates to the electron-positron vacuum polarization.
In this theory there is no habitual diagram for the vacuum
polarization with two vertices of the first order in $e$ because
operator $K_{1} $ is even.

When the external fermion line impulses lie on the mass surface
$\left( {p^{2} = m^{2}} \right)$, an interesting feature of the
theory is the compensation of the contribution of the fermion
propagator diagrams with that of the relevant terms in the
diagrams with vertices of the higher order of the expansion in
\textit{e}. Thus, in Fig. 3 the contribution of diagrams {\em a)}
and {\em c)} is compensated with the relevant parts of the
contribution of diagrams {\em b)} and {\em d)}; the contribution
of diagram {\em a)} in Fig. 4 is cancelled with the contribution
of the corresponding part of diagram {\em b)}; in Fig. 6 the
contribution of diagrams {\em a), b), c)} is cancelled with that
of the corresponding part of diagram {\em d)}, a similar
compensation occurs for diagrams {\em e), f), g)} and {\em h); i),
j), k)} and {\em l)}, respectively. For the real external fermions
under discussion the vertex operators $K_{n} $ get therewith
simplified significantly because of the law of conservation of
energy-momentum (see, e. g., (\ref{eq8}), (\ref{eq9})).

In view of the aforesaid, the scattering matrix can be expanded in
powers of \textit{e} so, that the matrix elements $S_{fi} $
contain no terms with electron-positron propagators. In so doing
the vertex operator matrix elements alter as follows:
\begin{eqnarray}
 && \label{eq12}\!\!\!\!\!\!\!\!\!\! \langle \vec {p}^{\,'}\left|
  {K_{1}}  \right|\vec {p}^{\,''} \rangle   \to   \langle \vec
{p}^{\,'}\left| {C} \right| \vec {p}^{\,''} \rangle;\\ &&
\label{eq13}\!\!\!\!\!\!\!\!\!\! \langle \vec {p}^{\,'}\left|
{K_{2}}  \right|\vec {p}^{\,''} \rangle   \to \sum\limits_{\nu
_{2} = \pm 1} {\int {d\vec {p}^{\,'''}\left\{ {\dfrac{{1}}{{\beta
E^{''} - \beta E^{'''} - \nu _{2}  k_{2}^{0}} } \langle \vec
{p}^{\,'}\left| {C} \right|\vec {p}^{\,'''} \rangle \langle \vec
{p}^{\,'''}\left| {C} \right|\vec {p}^{\,''} \rangle +} \right.}}
\nn\\ &&\!\!\!\!\!\!\!\!\!\! + \left.\dfrac{{1}}{{\beta E^{''} +
\beta E^{'''} - \nu _{2} k_{2}^{0}} } \langle \vec {p}^{\,'}\left|
{N} \right|\vec {p}^{\,'''} \rangle \langle \vec {p}^{\,'''}\left|
{N} \right|\vec {p}^{\,''} \rangle
 \right\};\\
&& \label{eq14}\!\!\!\!\!\!\!\!\!\!\langle \vec {p}^{\,'}\left|
{K_{3}}  \right|\vec {p}^{\,''} \rangle  \to \sum\limits_{\nu _{1}
\nu _{2} = \pm 1} {\int {d\vec {p}^{\,'''}d\vec
{p}^{\,IV}\dfrac{{1}}{{\left( {\beta E^{IV} - \beta E^{'} - \nu
_{1}  k_{1}^{0} - \nu _{2} k_{2}^{0}} \right) \left( {\beta
E^{'''} - \beta E^{'} - \nu _{1} k_{1}^{0}} \right)}}} \times} \nn
\\ &&\!\!\!\!\!\!\!\!\!\! \times \langle \vec {p}^{\,'}\left| {C}
\right|\vec {p}^{\,'''} \rangle \langle \vec {p}^{\,'''}\left| {C}
\right| \vec {p}^{\,IV}\!\! \rangle \langle \vec
{p}^{\,IV}\!\!\left| {C} \right|\vec {p}^{\,''} \rangle -
\dfrac{{1}}{{\left( {\beta E^{IV} + \beta E^{'} + \nu _{1}
k_{1}^{0} + \nu _{2} k_{2}^{0}} \right) \left( {\beta E^{'''} -
\beta E^{'} - \nu _{1} k_{1}^{0}} \right)}} \times  \nn\\ &&
\!\!\!\!\!\!\!\!\!\!\times  \langle \vec {p}^{\,'}\left| {C}
\right|\vec {p}^{\,'''} \rangle \langle \vec {p}^{\,'''}\left| {N}
\right| \vec {p}^{\,IV}\!\! \rangle \langle \vec
{p}^{\,IV}\!\!\left| {N} \right|\vec {p}^{\,''} \rangle +
\dfrac{{1}}{{\left( {\beta E^{IV} + \beta E^{'} + \nu _{1}
k_{1}^{0} + \nu _{2} k_{2}^{0}} \right) \left( {\beta E^{'''} +
\beta E^{'} + \nu _{1} k_{1}^{0}} \right)}} \times \nn\\ &&
\!\!\!\!\!\!\!\!\!\!\times \langle \vec {p}^{\,'}\left| {N}
\right|\vec {p}^{\,'''} \rangle \langle \vec {p}^{\,'''}\left| {C}
\right| \vec {p}^{\,IV}\!\! \rangle \langle \vec
{p}^{\,IV}\!\!\left| {N} \right|\vec {p}^{\,''} \rangle -
\dfrac{{1}}{{\left( {\beta E^{IV} - \beta E^{'} - \nu _{1}
k_{1}^{0} - \nu _{2}  k_{2}^{0}} \right) \left( {\beta E^{'''} +
\beta E^{'} + \nu _{1}  k_{1}^{0}} \right)}} \times\nn\\
&&\!\!\!\!\!\!\!\!\!\! \times \langle \vec {p}^{\,'}\left| {N}
\right|\vec {p}^{\,'''} \rangle \langle \vec {p}^{\,'''}\left| {N}
\right| \vec {p}^{\,IV}\!\! \rangle \langle \vec
{p}^{\,IV}\!\!\left| {C} \right|\vec {p}^{\,''} \rangle.
 \end{eqnarray}

Relations (\ref{eq13}), (\ref{eq14}) are close in their structure
to the formulas of the ``old'', noncovariant perturbation theory
developed by Heitler in the Dirac representation \cite{5}.
Expressions (\ref{eq13}), (\ref{eq14}) can be derived from the
relations of the Heitler's perturbation theory, if the \textit{FW}
transformation is performed for the free case of $A^{\mu} \left(
{x} \right) = 0$ in the matrix elements of the Heitler's
quantities $\left( {K_{n}}\right)_{Í}$ (therewith $\alpha _{\mu}
A^{\mu} \left( {x} \right)\, \to \,\left( {C + N} \right)$) and
nonzero terms (even in the upper and lower components of states $
\langle \left. {A_{FW}} \right|$ and $\left| {B_{FW} \rangle}
\right.$) are left in products $ \langle A_{FW} \left| {\left( {C
+ N} \right)\left( {C + N} \right)\ldots} \right|B_{FW} \rangle $
that have appeared. It seems that this rule can be extended to the
higher terms of the expansion in \textit{e} as well.

A significant difference of expressions (\ref{eq13}),
(\ref{eq14}), etc. from the formulas of ref. \cite{5} is absence
of any interaction between real electrons and positrons because
 the Hamiltonian in the \textit{FW} representation is even.
In this representation the electron-positron interaction can be
only between the real and intermediate virtual states.

\begin{figure}[p]
\noindent
\begin{minipage}[c]{75mm}
\includegraphics[width=75mm,height=55mm]{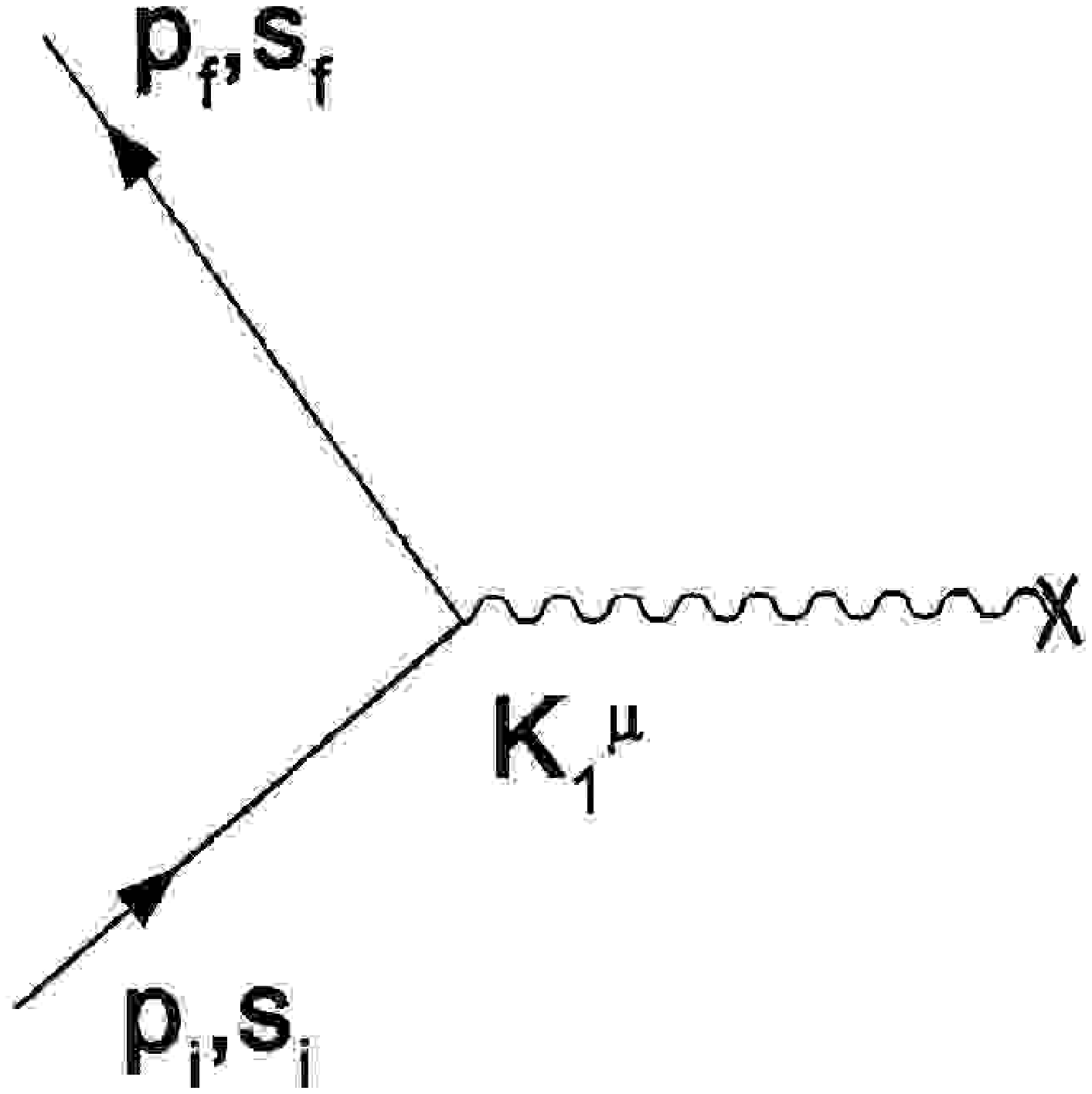}
\caption{Electron scattering in Coulomb field}
\end{minipage}\hfill\noindent
\begin{minipage}[c]{90mm}
\includegraphics[width=90mm,height=55mm]{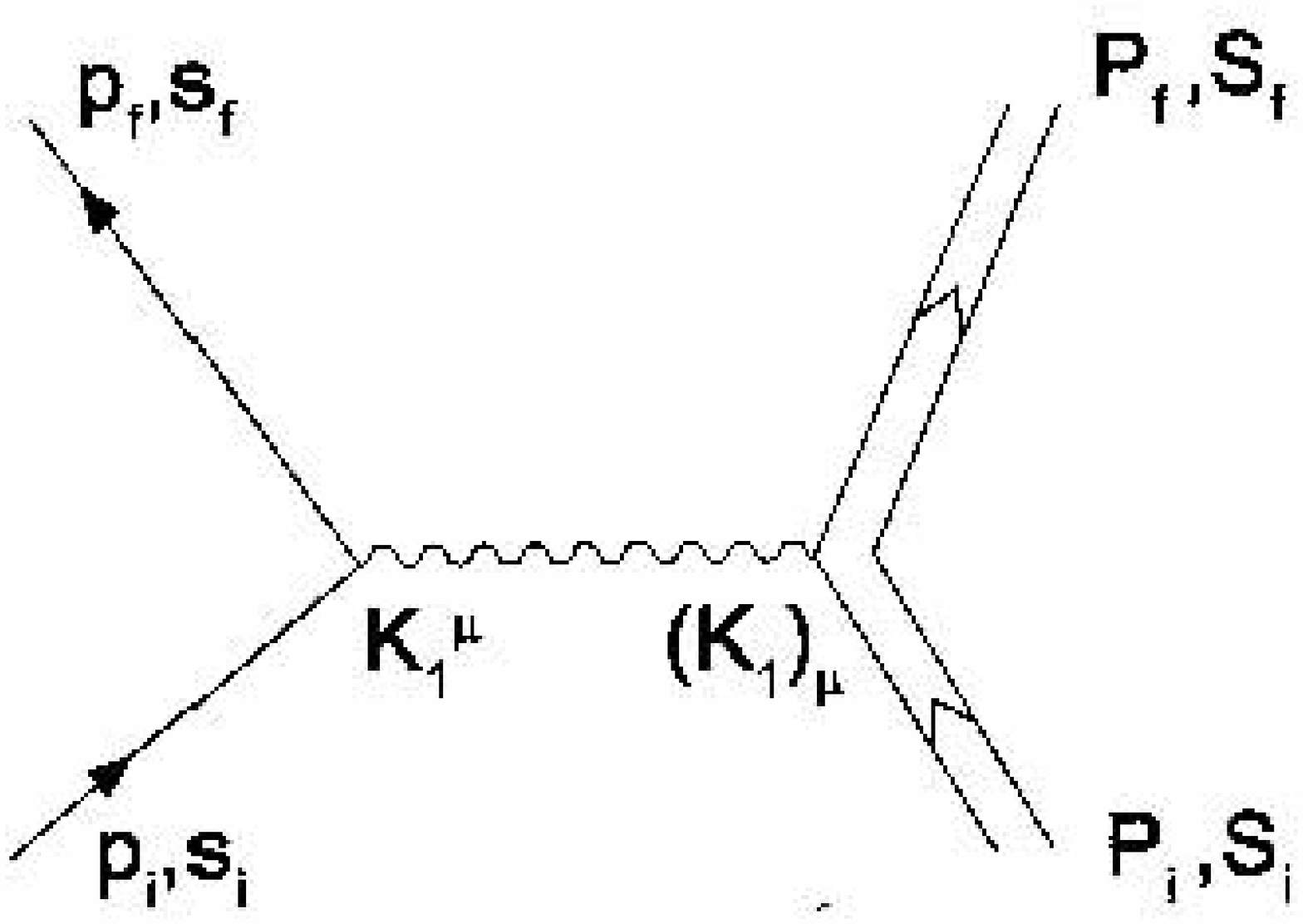}
\caption{Electron scattering on Dirac proton (M\"oller
scattering)}
\end{minipage}
\includegraphics[width=170mm,height=50mm]{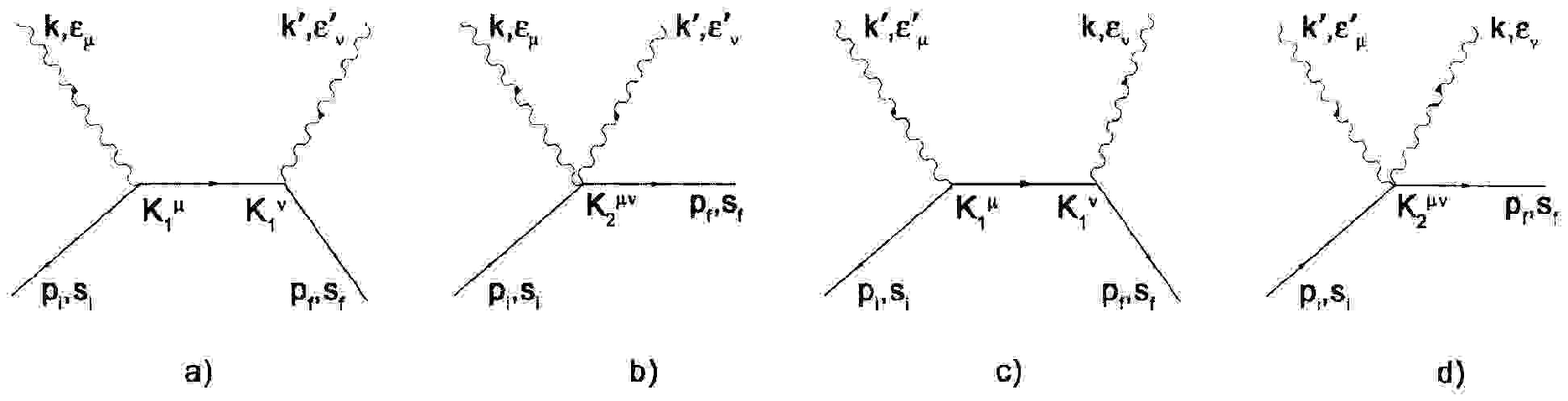}
\caption{Compton effect}
\includegraphics[width=170mm,height=35mm]{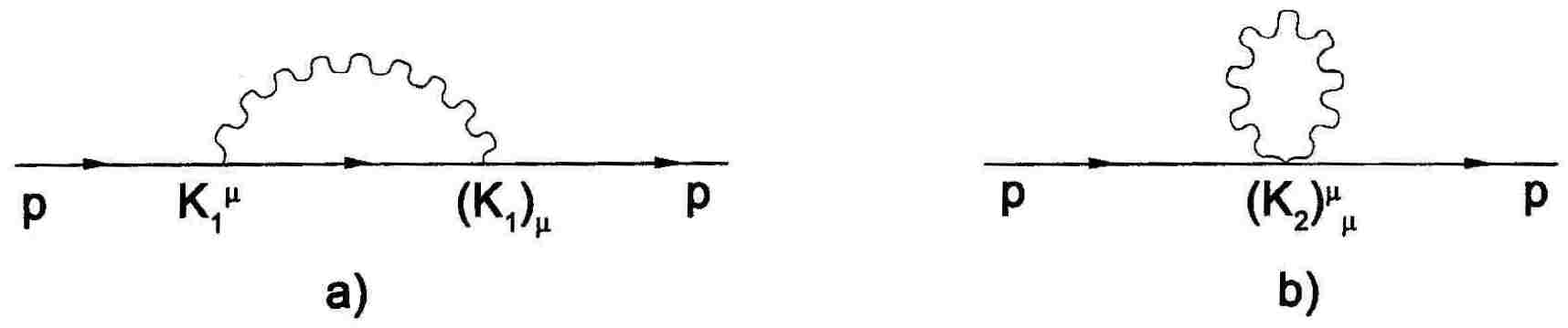}
\caption{Electron self-energy}
\begin{center}
\includegraphics[width=60mm,height=35mm]{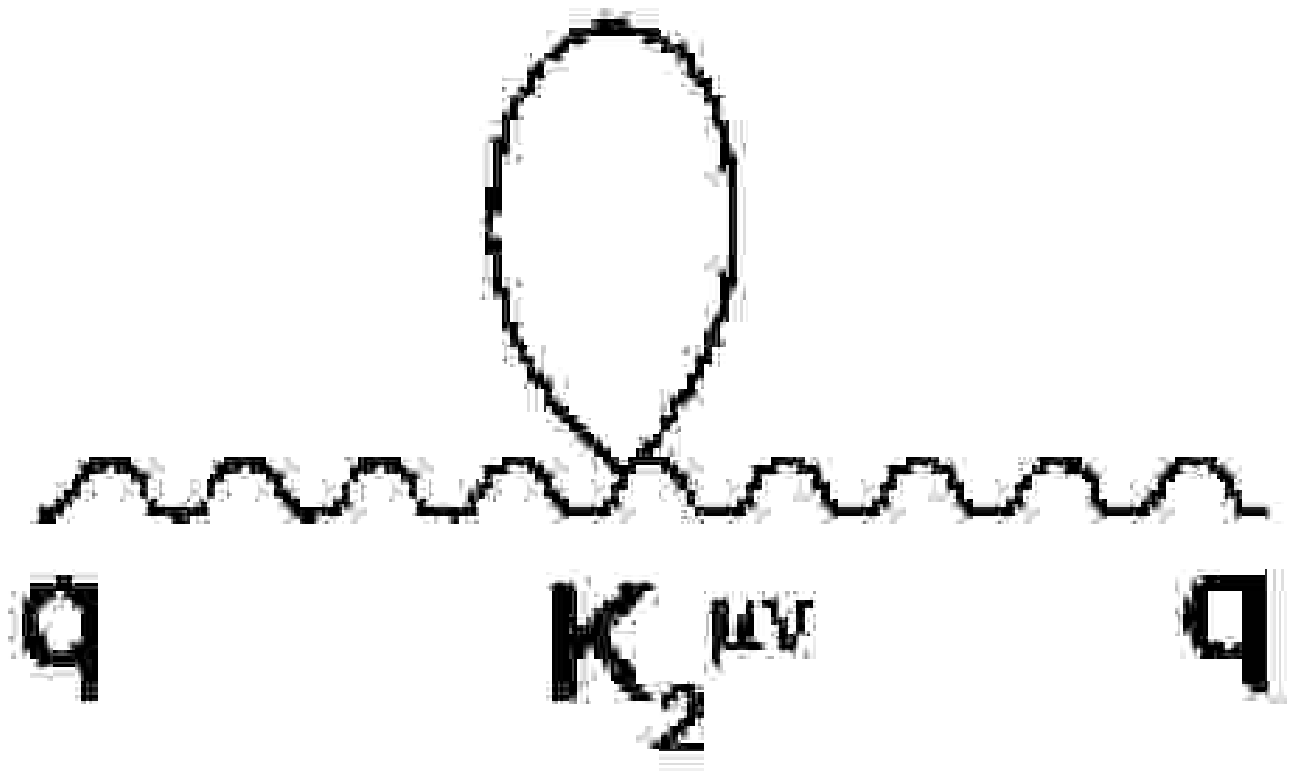}
\caption{Vacuum polarization}
\end{center}
\end{figure}
\begin{figure}[p]
\includegraphics[width=160mm,height=210mm]{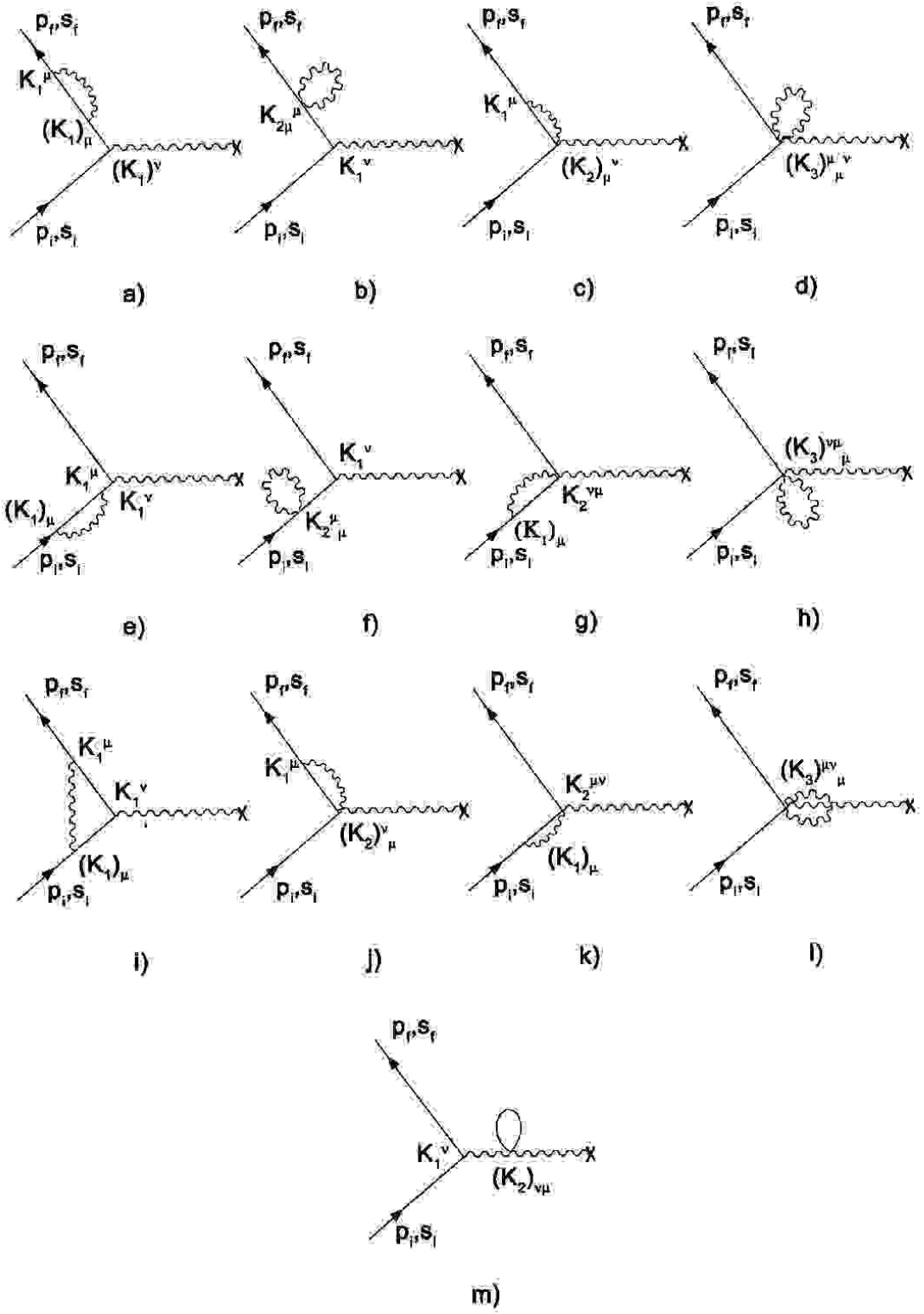}
\caption{Radiation corrections to the electron scattering in the
external field}
\end{figure}
\newpage
\subsection{Inclusion of electron-positron pair interactions\\ in the
$FW$ representation}

To construct real particle-antiparticle pair interaction processes
in the theory, additional terms must be introduced to Hamiltonian
\textit{H}$_{FW}$. A method to include processes with real
electron-positron pairs that would ensure proper results in the
calculation of QED effects (for example, electron-positron pair
annihilation cross section) is to introduce interaction between
positive-energy (negative-energy) states of equation (\ref{eq7})
and negative-energy (positive-energy) states of equation
(\ref{eq15}) derived by the \textit{FW} transformation of Dirac
equation (\ref{eq2}) with negative particle mass.
\begin{equation}
\label{eq15} p_{0}  \psi _{1FW} \left( {x} \right) = \left[ {\beta
E + K_{1} \left( { - m} \right) + K_{2} \left( { - m} \right) +
\ldots} \right]  \psi _{1FW} \left( {x} \right).
\end{equation}

Equation (\ref{eq7}) with the additional interaction can be
written as
\begin{equation} \label{eq16}
\begin{array}{l}
 p_{0}  \psi _{FW} \left( {x} \right) = \beta E  \psi _{FW}
\left( {x} \right) + \Bigg[ {K_{1} \left( {m,I,m} \right)}
 + \\+\left. \dfrac{{1}}{{2}}K_{2} \left( {m,I,m; m,I,m} \right) +
\dfrac{{1}}{{2}}K_{2} \left( {m,\gamma _{5} , - m; - m,\gamma _{5}
,m} \right) + \ldots \right]  \psi _{FW} \left( {x} \right) +
\\
 + \left[ K_{1} \left(  m,\gamma _{5} , - m \right) +
\dfrac{{1}}{{2}}K_{2} \left( {m,\gamma _{5} , - m;  - m,I, - m}
\right) + \right. \\ \left.  + \dfrac{{1}}{{2}}K_{2} \left(
{m,I,m; m,\gamma _{5} , - m} \right) + \ldots \right] \psi _{1FW}
\left( {x} \right)
 \end{array}
\end{equation}

In equation (\ref{eq16}), the notation of terms $K_{1},
K_{2},\ldots$ indicates the presence or absence of matrix $\gamma
_{5} $ near potentials $A^{\mu} \,\left( {x} \right)$ and the mass
sign on the left or on the right of fields $A^{\mu} \,\left( {x}
\right)$. Factor $\dfrac12$ of terms \textit{K}$_{2}$ is
introduced because of two possible ways of transition to the final
state of mass \textit{+m}. Similar to (\ref{eq16}), the additional
terms of interaction with field $\psi _{FW} \left( {x} \right)$
can be introduced to equation (\ref{eq15}) with negative mass.

The considered interaction between equations (\ref{eq7}),
(\ref{eq15}) can be formalized as follows. Introduce
eight-component field $\Phi _{FW} \left( {x} \right)$, in which
the four upper components are solutions to equation (\ref{eq7})
with positive mass $\left( { + m} \right)$ and the lower
components are solutions to equation (\ref{eq15}) with negative
mass $\left( { - m} \right)$. In the case under discussion
solutions (\ref{eq4}) for free motion are written as
$$\label{eq17a} \Phi _{FW}^{\left( { +}  \right)} \left( {x,s, +
m} \right) = \dfrac{{1}}{{\left( {2\pi}  \right)^{3/2}}}\left(
{\begin{array}{c}
 {U_{s}}  \\
 {0}
 \end{array}} \right)e^{-ip x};\eqno{(17a)}$$
$$\label{eq17b} \Phi _{FW}^{\left( { +}  \right)} \left( {x,s, -
m} \right) = \dfrac{{1}}{{\left( {2\pi}  \right)^{3/2}}}\left(
{\begin{array}{c}
 {0} \\
 {U_{s}}
 \end{array}} \right)e^{-ip x};\eqno{(17b)}$$
$$\label{eq17c} \Phi _{FW}^{\left( { -}  \right)} \left( {x,s, +
m} \right) = \dfrac{{1}}{{\left( {2\pi}  \right)^{3/2}}}\left(
{\begin{array}{c}
 {V_{s}}  \\
 {0}
 \end{array}}\right)e^{ip x};\eqno{(17c)}$$
$$\label{eq17d} \Phi _{FW}^{\left( { -}  \right)} \left( {x,s, -
m} \right) = \dfrac{{1}}{{\left( {2\pi}  \right)^{3/2}}}\left(
{\begin{array}{c}
 {0} \\
 {V_{s}}
 \end{array}} \right)e^{ip x}.\eqno{(17d)}$$

The extension of orthonormalization and completeness relations
(\ref{eq5}) to eight dimensions is quite evident.

Next, introduce matrices $8\times8$:
\begin{eqnarray*}
&& \beta _{1} = \left( \begin{array}{cc}
 I  & 0 \\
 0 &- I
 \end{array} \right),
\q\rho =\left(
\begin{array}{cc}
0 & \gamma _5 \\ \gamma _5 & 0
\end{array}
\right) , \\ &&\alpha ^k=\left(
\begin{array}{cc}
\alpha ^k & 0 \\ 0 & \alpha ^k
\end{array}
\right) ,\q \sigma ^k=\left(
\begin{array}{cc}
\sigma ^k & 0 \\ 0 & \sigma ^k
\end{array}
\right) ,\q \beta =\left(
\begin{array}{cc}
\beta & 0 \\ 0 & \beta
\end{array}
\right).\\
&&\beta_1^2=\rho^2=I;\q[\beta_1,\rho]_+=[\beta,\rho]_+=0;\\
&&[\beta_1,\alpha^k]_-=[\beta_1,\sigma^k]_-=[\beta_1,\beta]_-=0;\\
&&[\rho,\alpha^k]_-=[\rho,\sigma^k]_-=0.
\end{eqnarray*}
\setcounter{equation}{17}

Having substituted $m \to \beta _{1}  m$, equations (\ref{eq7}),
(\ref{eq15}) can be combined using the matrices $8\times8$ as
\begin{eqnarray}
\label{eq18} && p_{0} \Phi _{FW} \left( {x} \right) = \Big[\beta
E+K_1
  \left( {\beta _1 m,\left( {I+\rho}  \right),\beta _1 m} \right) +
 \nn\\
 && + \dfrac{1}{2}K_2 \left( {\beta _1 m ,\left( {I+\rho}  \right),
 \beta_1 m; \beta _1 m ,\left( {I+\rho}  \right) ,\beta _1 m}
\right) + \nn\\ &&\left. + \dfrac{1}{4}K_3 \left( {\beta _1 m_
,\left( {I+\rho}  \right) ,\beta _1 m; \beta _1 m ,\left( {I +
\rho} \right) ,\beta _1 m; \beta _1 m ,\left( {I+\rho}  \right)
,\beta _1 m} \right) + \ldots
 \right] \Phi_{FW}
 \left( {x} \right)
\end{eqnarray}

Equation (\ref{eq18}) contains equation (\ref{eq16}) for field
$\psi _{FW} \left( {x} \right)$ and relevant equation for field
$\psi _{1FW} \left( {x} \right)$. Interaction $\left( {I+\rho}
\right)  A^{\mu} \left( {x} \right)$ allows coupling between
solutions (17$a$) and (17$d$), (17$b$) and (17$c$), while there
is, as previously, no coupling between the other pairs of
solutions, (17$a$) and (17$c$), (17$b$) and (17$d$), as well as
(17$a$) and (17$b$), (17$c$) and (17$d$).

The calculation of the electron-positron pair annihilation cross
section in the second order of the perturbation theory with
inclusion of the introduced coupling between the solutions to
equations (\ref{eq7}) and (\ref{eq15}) is presented in Appendix 2.

The analysis shows that coupling $\left( {I+\rho}  \right) A^{\mu
}$, along with inclusion of interactions of real electron-positron
pairs, does not affect the physical QED results discussed in
Section 3. Processes with real negative-mass fermions appear in
the theory.

Thus, with the coupling $\left( {I+\rho}  \right)  A^{\mu} $, the
theory in the $FW$ representation is symmetric about the particle
(antiparticle) mass sign, however, the signs in masses of the
particle and antiparticle interacting with each other should be
opposite.

In the theory of processes with real particle-antiparticle pairs,
another possible method of inclusion is to introduce the coupling
between equations of motion for electron and positron in field
$A^{\mu} \left( {x} \right)$. The analysis of the coupling
introduction is not discussed in this review paper.

\subsection{$C$, $P$, $T$-symmetries in the
$FW$ representation}

The Hamiltonian of equation (\ref{eq18}) is invariant under
spatial reflections $\vec {x} \to \vec {x}^{'} = - \vec {x}$.
Hence, the solutions to equation (\ref{eq18}) are preserved in
$P$-inversion with an accuracy to phase factor and matrices
commutating with operator $p_{0} $ and generalized Hamiltonian
\textit{H}$_{FW} $.

Consider two cases:
\begin{equation}\label{eq19}
 1.\q\Phi _{FW}^{'} \left(
{\vec {x}^{'},t} \right) = P  \Phi _{FW}^{} \left( {\vec {x},t}
\right) = e^{i\varphi} \beta \Phi _{FW}^{} \left( {\vec {x},t}
\right).\qq\qq\qq\qq\qq\qq\qq
\end{equation}
In this case, like in the Dirac representation, the particles and
antiparticles have opposite internal parity, which corresponds to
the existing experimental data.

 $$\!\!\!\!\!\!\!\! 2. \q\Phi _{FW}^{'}
\left( {\vec {x}^{'},t} \right) = P  \Phi _{FW}^{} \left( {\vec
{x},t} \right) = e^{i\varphi} \Phi _{FW}^{} \left( {\vec {x}^{},t}
\right).\qq\qq\qq\qq\qq\qq\qq$$ For this case, which is also
admitted by this theory, the particles and antiparticles in the
$FW$representation have the same internal parity.

In $C$-conjugation equation (\ref{eq18}) transforms to the
equation for charge-conjugate spinor $\Phi _{FW}^{C} $ with
changed charge and mass signs.
\begin{eqnarray}
\label{eq20} && P_{0} \Phi _{FW}^{C} \left( {x} \right) = \left[
{\beta E - K_{1} \left( { - \beta _{1} m; A^{0}; A^{k}} \right) +}
\right. \nn\\ && + K_{2} \left( { - \beta _{1} m; A^{0}; A^{k}}
\right) - K_{3} \left( { - \beta _{1} m; A^{0}; A^{k}} \right) +
\ldots\left. {} \right]\Phi _{FW}^{C} \left( {x} \right).
 \end{eqnarray}

It is easy to show using (\ref{eq8}), (\ref{eq9}), (\ref{eq18})
that the theory is $C$-invariant, if
\begin{equation}
\label{eq21} \Phi _{FW}^{C} \left( {x} \right) = C  \Phi
_{FW}^{\ast}  \left( {x} \right) = \rho \sigma ^{\left( {2}
\right)}\Phi _{FW}^{\ast}  \left( {x} \right).
\end{equation}

Now consider time inversion $t \to {t}' = - t$. Equation
(\ref{eq18}) transforms to the equation for function $\Phi
_{FW}^{'} \left( {t^{'}} \right)$ with a changed sign in potential
vectors $A^{k}$
\begin{equation}
\label{eq22} P_{0} ^{\prime} {\Phi} '_{FW} \left( {t^{'}} \right)
= \left[ {\beta E + K_{1} \left( {\beta _{1} m; A^{0};  - A^{k}}
\right) + K_{2} \left( {\beta _{1} m; A^{0};  - A^{k}} \right) +
\ldots} \right] \Phi _{FW}^{'} \left( {t^{'}} \right).
\end{equation}

It can be shown that for the $T$-invariance it is necessary that
\begin{equation}
\label{eq23} \Phi _{FW}^{'} \left( {t^{'}} \right) = T\Phi
_{FW}^{\ast}  \left( {t^{}} \right) = \sigma ^{\left( {2}
\right)}\Phi _{FW}^{\ast}  \left( {t^{}} \right).
\end{equation}

Relations (\ref{eq19})---(\ref{eq23}) suggest that, like in the
Dirac representation, quantum electrodynamics in the
$FW$representation is invariant under the $P,C,T$-transformations.

\section{WEAK INTERACTION\\ IN THE $FW$ REPRESENTATION}

For illustration this paper considers the weak interaction in the
$V$-$A$ form of current-current four-fermion interaction in the
lower order of the perturbation theory. The study of features of
the electroweak theory in the $FW$ representation will be a
subject of the following publications.

As seen from the M\"oller scattering consideration (diagram in
Fig.~2) with taking into account (\ref{eq12}), for the external
lines corresponding to real fermions the electromagnetic vector
current in the \textit{FW} representation, in the first order in
\textit{e}, is
\begin{equation}
\label{eq24} \left( {j_{FW}^{V}}  \right)^{\mu}  = \Phi _{FW}^{ +}
\left( {x} \right) C_{el}^{\mu}   \Phi _{FW} \left( {x} \right)
\end{equation}

In (\ref{eq24}), according to (\ref{eq18}), $C_{el}^{\mu}  $ is of
structure $C_{el}^{\mu} \,\left( {\beta _{1} m,\,\left( {I+\rho}
\right),\,\beta _{1} m} \right)$. Thus, with the accuracy to the
first order in \textit{e},
\[
\left( {j_{D}^{V}}  \right)^{\mu}  = \psi _{D}^{ +}  \left( {x}
\right) \alpha ^{\mu} \psi _{D} \left( {x} \right) \to \left(
{j_{FW}^{V}} \right)^{\mu}  = \Phi _{FW}^{ +}  \left( {x} \right)
C_{el}^{\mu}  \Phi _{FW} \left( {x} \right).
\]

Similarly, axial current in the \textit{FW} representation can be
obtained with the same accuracy using the methods of ref.
\cite{4}:
\begin{equation}
\label{eq25} \left( {j_{D}^{A}}  \right)^{\mu}  = \psi _{D}^{ +}
\left( {x} \right)\alpha ^{\mu}   \gamma ^{5}\psi _{D} \left( {x}
\right) \to \left( {j_{FW}^{A}}  \right)^{\mu}  = \Phi _{FW}^{ +}
\left( {x} \right) \left( {N_{el}^{'}} \right)^{\mu} \Phi _{FW}
\left( {x} \right).
\end{equation}

In (\ref{eq25}) $\left( {N_{el}^{'}}  \right)^{\mu} $ differs from
the previously introduced vector $N^{\mu} $ in the extension to
eight dimensions and the $8\times8$ matrix
 $\gamma _{5} =
\left( {\begin{array}{cc}
 \gamma _{5}  &0 \\
 0 &\gamma _{5}
 \end{array}} \right)$, located near the matrix
$\left( {I+\rho}  \right)$.
\begin{equation}
\label{eq26}
\begin{array}{l}
 \left( {N_{el}^{'}}  \right)^{0} = eR\left( {L  \left( {I+\rho}
\right)  \gamma _{5} - \left( {I+\rho}  \right)  \gamma _{5}^{}
 L} \right)  R; \\
 \left( {N_{el}^{'}}  \right)^{k} = - eR\left( {\alpha ^{k}\left( {I +
\rho}  \right)  \gamma _{5} - L  \alpha ^{k}\left( {I+\rho}
\right)  \gamma _{5}^{}  L} \right)  R.
 \end{array}
\end{equation}

In view of (\ref{eq24}), (\ref{eq25}), weak \textit{V-A} current
can be written as
\begin{equation}
\label{eq27} \left( {j_{FW}^{weak}}  \right)^{\mu}  = \Phi _{FW}^{
+} \left( {x} \right) \left( {C_{weak}^{\mu}  - \left(
{N_{weak}^{'}} \right)^{\mu} } \right)\Phi _{FW} \left( {x}
\right)
\end{equation}
 with substitution $C_{el}^{\mu}  $ and $\left( {N_{el}^{'}}
\right)^{\mu} \quad e^{2} \to \dfrac{{G}}{{\sqrt {2}} }$ where
\textit{G} is the Fermi constant of weak interaction.

The amplitudes of weak-interaction processes in the $FW$
representation are
\begin{equation}
\label{eq28} M = \left( {j_{FW}^{weak}}  \right)^{\mu}  \left(
{j_{FW}^{weak}} \right)_{\mu} ^{ +}.
\end{equation}

The amplitude $M$ can be shown to be nonvariant taken separately
under $P$ and $C$ reflections, but invariant under the combined
$CP$-inversion.

Expressions (\ref{eq27}), (\ref{eq28}) allow us to calculate
amplitudes of specific weak-interaction processes in the
Foldy-Wouthuysen representation in the lower order of the
perturbation theory. The final results of the calculations agree
with those in the Dirac representation.

The amplitude of weak processes in the $FW$ representation in form
(\ref{eq28}) are degenerate relative to mass signs in particles
and antiparticles. Particles of mass $ + m$ interact with each
other and antiparticles of mass $ - m$. Conversely, particles of
mass $ - m$ interact with each other and with antiparticles of
mass $ + m$.

An interesting feature of the theory is the possibility to
associate the break of $CP$-invariance with total or partial
removal of the degeneracy in the particle (antiparticle) mass
sign.

In fact, if $I+\rho $ is substituted for $\dfrac{{1}}{{2}}\left(
{I+\beta _{1}}  \right)\left( {I+\rho}  \right)$ in terms
$C_{weak}^{\mu}  ,\left( {{N}'_{weak}}  \right)^{\mu} $ in
expression for weak current (\ref{eq27}), then, on the one hand,
the $CP$-invariance of the theory is broken because of
anticommutation of matrices $\beta _{1} ,\rho $ and, on the other
hand, operator $\dfrac{{1}}{{2}}\left( {I+\beta _{1}} \right)$
chooses solutions with particle and antiparticle mass $ + m$ from
(17), which makes the theory asymmetric about the mass sign. The
degeneracy in mass sign can be removed completely, if difference
$C_{weak}^{\mu}  - \left( {{N}'_{weak}} \right)^{\mu} $ in
(\ref{eq27}) is multiplied, for example, from the left by
$\dfrac{{1}}{{2}}\left( {I+\beta}  \right) \dfrac{{1}}{{2}}\left(
{I+\beta _{1}}  \right)$. In this case the theory is also
$CP$-nonvariant, and the only solution (17$a$) with particle mass
$ + m$ is chosen from (17). In this case particles of mass $ + m$
can interact with one another and, due to multiplier $I+\rho $ in
$C_{weak}^{\mu}  $ and $\left( {{N}'_{weak} } \right)^{\mu} $,
with antiparticles of mass $\left( { - m} \right)$. Upon the
$CP$-inversion, conversely, only particles of mass $ - m$ can
interact with one another and there is a possibility of the
interaction with antiparticles of mass $ + m$.

The theory admits incomplete removal of the degeneracy in mass
sign, if $\dfrac{{1}}{{2}}\left( {I+\beta}  \right)\times\lb\times
\dfrac{{1}}{{2}}\left( {I+\beta _{1}}  \right)$ is replaced by
$\dfrac{{1}}{{2}}\left( {I + \varepsilon \beta}  \right)
\dfrac{{1}}{{2}}\left( {I + \varepsilon _{1} \beta _{1} }
\right)$, where $|\varepsilon|  \le 1$, $|\varepsilon _{1}|  \le
1$. The $CP$-invariance break extent depends on $\varepsilon$,
$\varepsilon _{1} $.

\section*{CONCLUSION}

From the results of this review it follows that the interacting
field theory version under discussion in the $FW$ representation
has some new physical consequences in comparison with the Dirac
representation. When the interaction of real particles with
antiparticles is included, negative-mass (but positive-energy)
fermions appear in the theory. The theory is symmetric about the
mass sign, but the particle and antiparticle masses should be of
opposite signs. In the theory there is a possibility to associate
the break of $CP$-symmetry with the break of symmetry in particle
(antiparticle) mass sign.

The above consequences foster further theoretical studies of the
Foldy-Wouthuysen representation and comparison of their results to
experimental data. In particular, it is necessary to consider
electroweak theory and quantum chromodynamics in the $FW$
representation to detect and analyze new resulting physical
consequences.

\newpage
\setcounter{equation}{0}
\renewcommand{\theequation}{A.\arabic{equation}}
\begin{flushright}
{\bf APPENDIX 1}
\end{flushright}

\section*{DERIVATION OF HAMILTONIAN $H_{FW}$\\ OF DIRAC
 EQUATION IN THE\\
FOLDY-WOUTHUYSEN REPRESENTATION}

In notation of ref. \cite{4},
\begin{equation}\label{S1}
   H_{FW} = U_{FW} H_{D} U_{FW}^{ +}  - iU_{FW}  \dfrac{{\partial
}}{{\partial t}}\left( {U_{FW}^{ +} }  \right).
\end{equation}

In (A.1) $U_{FW} = \left( {1 + \delta _{1} + \delta _{2} + \delta
_{3} + \ldots} \right)  U_{0} ,$
 $\delta _{1} \sim e,\quad \delta _{2} \sim e^{2},\quad \delta _{3}\sim e^{3},\quad
U_{0} = R\left( {1 + L} \right)$ is the Foldy-Wouthuysen
transformation matrix at $A^{\mu} \left( {x} \right) = 0$. For
Dirac Hamiltonian of free motion $\left( {H_{0}}  \right)_{D} $,
$U_{0} \left( {H_{0}}  \right)_{D} U_{0}^{ + } = \beta E$ is
valid.

From the condition of unitarity $U_{FW}^{ +}   U_{FW} = 1$ it
follows that
 \begin{equation} \left\{ \begin{array}{l}
 {\delta _{1}^{ +}  = - \delta _{1}};  \\
 {\delta _{2}^{ +}  = - \delta _{2} + \delta _{1}  \delta _{1}};  \\
 {\delta _{3}^{ +}  = - \delta _{3} + \delta _{2}  \delta _{1} + \delta
_{1}  \delta _{2} - \delta _{1}  \delta _{1}  \delta _{1}};
\\
 {..................................................................}
 \end{array}  \right.
\end{equation}

In view of (A.2),
\begin{equation}
 \begin{array}{l}
 H_{FW} = \beta E + K_{1} + K_{2} + K_{3} + \ldots, \\
 K_{1} = \delta _{1}  \beta E - \beta E  \delta _{1} +
i\dfrac{{\partial \delta _{1}} }{{\partial t}} + C + N, \\[+10pt]
 K_{2} = \delta _{2}  \beta E - \beta E  \delta _{2} +
i\dfrac{{\partial \delta _{2}} }{{\partial t}} - K_{1} \delta _{1}
+ \delta _{1} \left( {C + N} \right), \\[+10pt]
 K_{3} = \delta _{3}  \beta E - \beta E  \delta _{3} +
i\dfrac{{\partial \delta _{3}} }{{\partial t}} - K_{1} \delta _{2}
- K_{2}  \delta _{1} + \delta _{2} \left( {C + N} \right).
 \end{array}
 \end{equation}

Introduce the notions of even (with superscript $e$) and odd (with
superscript $o$) operators that do not relate and relate,
respectively, the upper and lower wave function components. Ref.
\cite{4} establishes the following relation between the even and
odd operators $\delta _{i} $.
 \begin{equation}
\begin{array}{l}
 \delta _{1}^{e}  R + R  \delta _{1}^{e} = R  L  \delta
_{1}^{o} - \delta _{1}^{o}  L  R; \\
 \delta _{2}^{e}  R + R  \delta _{2}^{e} = R  L  \delta
_{2}^{o} - \delta _{2}^{o}  L  R - RL\left( {\delta _{1} \delta
_{1}}  \right)^{o} + R\left( {\delta _{1}  \delta _{1}}
\right)^{e}; \\
 \delta _{3}^{e}  R + R  \delta _{3}^{e} = R  L  \delta
_{3}^{o} - \delta _{3}^{o}  L  R - RL \left( {\delta _{2} \delta
_{1} + \delta _{1}  \delta _{2} - \delta _{1}  \delta _{1} \delta
_{1}}  \right)^{o} + \\
 + R\left( {\delta _{2}  \delta _{1} + \delta _{1}  \delta _{2} -
\delta _{1}  \delta _{1}  \delta _{1}}  \right)^{e}.
 \end{array}
 \end{equation}

As by definition $K_{1} , K_{2} , K_{3} \ldots$ are even
operators, in (A.3) the odd terms must be set equal to zero,
  \begin{equation}
 \left\{ \begin{array}{l}
 {\delta _{1}^{o}  \beta E - \beta E  \delta _{1}^{o} +
i\dfrac{{\partial \delta _{1}^{o}} }{{\partial t}} + N = 0};
\\[+10pt]
 {\delta _{2}^{o}  \beta E - \beta E  \delta _{2}^{o} +
i\dfrac{{\partial \delta _{2}^{o}} }{{\partial t}} - K_{1} \delta
_{1}^{o} + \delta _{1}^{o}  C + \delta _{1}^{e}  N = 0};
\\[+10pt]
 {\delta _{3}^{o}  \beta E - \beta E  \delta _{3}^{o} +
i\dfrac{{\partial \delta _{3}^{o}} }{{\partial t}} - K_{1} \delta
_{2}^{o} - K_{2}  \delta _{1}^{o} + \delta _{2}^{o}  C + \delta
_{2}^{e}  N = 0}; \\

{.........................................................................................................}
 \end{array}  \right.
  \end{equation}

Then the Hamiltonian expansion terms $K_{1} ,K_{2} , K_{3} \ldots$
are determined as
\begin{equation}\left\{ {\begin{array}{l}
 {K_{1} = \delta _{1}^{e} \beta E - \beta E  \delta _{1}^{e} +
i\dfrac{{\partial \delta _{1}^{e}} }{{\partial t}} + C} ;\\[+10pt]
 {K_{2} = \delta _{2}^{e} \beta E - \beta E  \delta _{2}^{e} +
i\dfrac{{\partial \delta _{2}^{e}} }{{\partial t}} - K_{1} \delta
_{1}^{e} + \delta _{1}^{e}  C + \delta _{1}^{o}  N};
\\[+10pt]
 {K_{3} = \delta _{3}^{e} \beta E - \beta E  \delta _{3}^{e} +
i\dfrac{{\partial \delta _{3}^{e}} }{{\partial t}} - K_{1} \delta
_{2}^{e} - K_{2}  \delta _{1}^{e} + \delta _{2}^{e}  C + \delta
_{2}^{o}  N}. \\

{..............................................................................................................}
 \end{array}}  \right.
 \end{equation}

Operator equalities (A.6) simultaneously with (A.4), (A.5) allow
us to determine Hamiltonian $H_{FW} $ as a series in terms of
powers of $e$.

For the external lines corresponding to real electrons
(positrons), the first three terms in (A.6) vanish in the matrix
elements due to the law of conservation of energy. In this case,
in the impulse representation, the matrix elements $K_{1} , K_{2}
,K_{3} \ldots$ are determined by relations (\ref{eq12}),
(\ref{eq13}), (\ref{eq14}).
\bigskip
\begin{flushright}
{\bf APPENDIX 2}
\end{flushright}
\section*{CALCULATIONS OF QED PROCESSES\\ IN THE $FW$ REPRESENTATION}

\subsection*{1\quad Electron scattering in Coulomb field  ${A_{0} \left(
{x} \right) = - \dfrac{{Ze}}{{4\pi \left| \vec x \right|}}}$}
\begin{eqnarray*} && S_{fi} = - i\int d^{4}x\Psi
_{FW}^{\left( { +}  \right)} \left( {x,p_{f} ,s_{f}}
\right)K_{1}^{0}  A_{0} \Psi _{FW}^{\left( { +} \right)} \left(
{x,p_{i} ,s_{i}}  \right) = \\ && = - \dfrac{{i\delta \left(
{E_{f} - E_{i}}  \right)}}{{\left( {2\pi}
\right)^{2}}}U_{s_{_{\!\!f}}} ^{ +}  \langle \vec {p}_{f} \left|
{C} \right.^{0}A_{0} \left| {\vec {p}_{i}} \right. \rangle
U_{s_{i}} =
\\ && = i\dfrac{{Ze^{2}}}{{\vec {q}^{\,2}}}\dfrac{{\delta \left(
{E_{f} - E_{i}} \right)}}{{\left( {2\pi}
\right)^{2}}}U_{s_{_{\!\!f}}} ^{ +} \dfrac{{1}}{{2E_{i} }}\left(
{E_{i} + m + \dfrac{{1}}{{E_{i} + m}}\vec {\sigma} } \vec {p}_{f}
\vec {\sigma}  \vec {p}_{i} \right)U_{s_{i}}  , \\
 &&\vec {q} = \vec {p}_{f} - \vec {p}_{i} .
 \end{eqnarray*}

Notation $K_{1}^{0}  A_{0} $ made for convenience actually means
$K_{1}^{0}  A_{0} \equiv K_{1} $ at $\vec {A}\left( {x} \right) =
0$. That is, $A_{0} \left( {x} \right)$ gets into the places
determined by expression (\ref{eq8}). The same is true for
notation $C^{0}  A_{0} $. The transition from $K_{1}^{0} A_{0} $
to $C^{0}  A_{0} $ is performed in accordance with (\ref{eq12}).

Then ordinary methods in conjunction with matrix element $S_{fi} $
can be used to obtain Mott differential scattering cross section
that transfers to Rutherford one in the nonrelativistic case.

\subsection*{2\quad Electron scattering on Dirac proton (M\"oller scattering)}

\begin{eqnarray*}
&& S_{fi} = - i\int d^{4}xd^{4}y\Psi _{FW}^{\left(  + \right)}
\left( {x,p_{f} ,s_{f}}  \right)K_{1}^{\alpha}   \Psi
_{FW}^{\left( { +} \right)} \left( {x,p_{i} ,s_{i}}  \right) D_{F}
\left( {x - y} \right)  \times\\ && \times \Psi _{FW}^{\left( { +}
\right)} \left( {y,P_{f} ,S_{f}} \right)\left( { - K_{1}}
\right)_{\alpha}   \Psi _{FW}^{\left( { +} \right)} \left(
{y,P_{i} ,S_{i}}  \right) = \\ && = - \dfrac {i\delta ^{4}\left(
{P_{f} - P_{i} + p_{f} - p_{i}} \right)} {\left( {p_{f} - p_{i}}
\right)^{2}} 2\pi \left( {U_{s_{_{\!\!f}}} ^{ + } \langle \vec
{p}_{f} \left| {C^{\alpha} }\right| {\vec {p}_{i}}  \rangle
U_{s_{i}} }  \right)\left( {U_{S_{f}} \langle \vec {P}_{f} \left|
{C_{\alpha }} \right| {\vec {P}}_{i} \rangle U_{S_{i}} } \right)
\end{eqnarray*}
Matrix element $S_{fi}$ determines M\"oller electron scattering
cross section.
\subsection*{3\quad Compton effect}
\begin{eqnarray*}
&& S_{fi} = - i\left( {\chi _{s_{_{\!\!f}}} ^{ +}\quad  0}
\right)\left\{
 {\int} \dfrac{{d^{4}zd^{4}yd^{4}p_{1} ^{_{}}
}}{{\left( {2\pi} \right)^{7}\sqrt {2k_{0}^{'}  2k_{0}} } }\left(
{e^{ip_{f} y}  (K_{1})_\mu   \varepsilon ^{'\mu}
e^{ik^{'}y}\dfrac{{e^{ - ip_{1}  y}}}{{p_{1}^{0} - \beta E\left(
{\vec {p}_{1}}  \right)}}  e^{ip_{1} z} (K_{1})_{\nu}
\times}\right.\right.
\\
 &&\left.\times \varepsilon ^{\nu}   e^{ - ikz}  e^{ip_{i} z} +
e^{ip_{f}  y}  (K_{1})_\mu   \varepsilon ^{\mu}  e^{ -
iky}\dfrac{{e^{ - ip_{1} y}}}{{p_{1}^{0} - \beta E\left( {\vec
{p}_{1}} \right)}}e^{ip_{1}  z}  (K_{1})_\nu {\varepsilon}^{'\nu}
 e^{ik'z}  e^{ - ip_{i} z} \right) +
\\
 &&+ \int d^{4}y\dfrac{{1}}{{\left( {2\pi}  \right)^{3}\sqrt {2k_{0}^{'}
 2k_{0}} } }\left( {e^{ip_{f} y}  (K_{2})_{\mu \nu}
 \varepsilon ^{'\mu}   e^{ik'y}  \varepsilon ^{\nu}
 e^{ - iky}  e^{ - ip_{i}  y} + }\right.\\
 &&\left.+ e^{ip_{f}  y}  (K_{2})_{\mu \nu}  \varepsilon ^{\mu}   e^{ -
iky}  {\varepsilon} ^{'\nu}  e^{ik'y}\right) \Bigg\} \left(
{{\begin{array}{c}
 {\chi_{s_{i}} } \\
 {0}
\end{array}} } \right).
 \end{eqnarray*}

The first integral combines the contribution made by diagrams {\em
a)} and {\em c)} in Fig.~3, the second one combines that made by
diagrams {\em b)} and {\em d)}.

By record $(K_{1})_\mu   \varepsilon ^{\mu}   e^{ - iky}$,
$(K_{2})_{\mu \nu}  \varepsilon ^{\mu}   e^{ - iky} {\varepsilon
}^{'\nu}   e^{ik'y}$, etc. is meant the same as specified in item
1 of Appendix 2.

From the law of conservation of energy-momentum, the contribution
of the terms in the first square bracket of expression (\ref{eq9})
for $K_{2} $ to matrix element $S_{fi} $ is zero. For the same
reason the contribution of the first two terms in the second
square bracket vanishes. Then, the third and fourth terms in the
second square bracket are compensated by the terms in the first
integral in the expression for $S_{fi} $ that correspond to the
contribution of diagrams $a$) and $c$) in Fig.~3. Thus, a
contribution to the matrix element $S_{fi} $ is made only by the
last four terms in expression (\ref{eq9}). The above facts are
general in calculations of the second-order processes of the
perturbation theory with impulses of electron lines lying on the
mass surface. In view of the aforesaid,
 $S_{fi} = \dfrac{{ - i\left( {2\pi}  \right)^{4}  \delta ^{4}\left(
{p_{i} + k - p_{f} - k'} \right)}}{{\left( {2\pi} \right)^{3}\sqrt
{2k_{0}  2k_{0}^{'}} } }\left( {\chi _{s_{_{\!\!f}}} ^{ +}\quad 0}
\right)  A\left( {{\begin{array}{c}
 {\chi_{s_{i}} } \\
 {0}
\end{array}} } \right),\quad $ where
\begin{eqnarray*}
&& A = C_{\mu}  \left( {\vec {p}_{f} ; \vec {p}_{i} + \vec {k}}
\right)  \varepsilon ^{'\mu} \dfrac{{\dfrac{{1}}{{2}}\left( {I +
\beta} \right)}}{{\beta  E\left( {\vec {p}_{i}}  \right) + k_{0} -
E\left( {\vec {p}_{i} + \vec {k}} \right)}}C_{\mu} \left( {\vec
{p}_{i} + \vec {k}; \vec {p}_{i}}  \right)  \varepsilon ^{\mu}  +
\\
 && + N_{\mu}   {\left( {\vec
{p}_{f} ; \vec {p}_{i} + \vec {k}} \right)  \varepsilon ^{'\mu}
\dfrac{{\dfrac{{1}}{{2}}\left( {I-\beta} \right)}}{{ - \beta
E\left( {\vec {p}_{i}}  \right) + k_{0} + E\left( {\vec {p}_{i} +
\vec {k}} \right)}}N_{\mu}  \left( {\vec {p}_{i} + \vec {k}; \vec
{p}_{i}}  \right)  \varepsilon ^{\mu}  + }\\ && + C_{\mu} \left(
{\vec {p}_{f} ; \vec {p}_{i} - {\vec {k}}'} \right) \varepsilon
^{\mu} \dfrac{{\dfrac{{1}}{{2}}\left( {I+\beta} \right)}}{{\beta
 E\left( {\vec {p}_{i}}  \right) - {k}'_{0} - E\left( {\vec
{p}_{i} - {\vec {k}}'} \right)}}C_{\mu} \left( {\vec {p}_{i} -
{\vec {k}}'; \vec {p}_{i}} \right)  {\varepsilon} ^{'\mu}  +
\\ && + N_{\mu}   \left( {\vec {p}_{f} ; \vec {p}_{i} - \vec {k}'}
\right)  \varepsilon ^{\mu} \dfrac{{\dfrac{{1}}{{2}}\left(
{I-\beta} \right)}}{{ - \beta E\left( {\vec {p}_{i}}  \right) -
{k}'_{0} + E\left( {\vec {p}_{i} - {\vec {k}}'} \right)}}N_{\mu}
\left( {\vec {p}_{i} - {\vec {k}}'; \vec {p}_{i}}  \right)
{\varepsilon} ^{'\mu}
 \end{eqnarray*}

\begin{eqnarray*}
&&\!\!\!\!\!\!\!\!\!\! C^{\mu} \left( {\vec {p}_{f} ; \vec {p}_{i}
+ \vec {k}} \right) = \left\{ \begin{array}{l}
 {e  R_{f}  R_{1} \left( {1 + \dfrac{{\vec {\alpha} \vec {p}_{f}
}}{{E\left( {\vec {p}_{f}}  \right) + m}}  \dfrac{{\vec {\alpha}
\left( {\vec {p}_{i} + \vec {k}} \right)}}{{E\left( {\vec {p}_{i}
+ \vec {k}} \right) + m}}} \right),\quad \mu = 0}; \\
 { - e  R_{f}  R_{1}  \beta \left( {\dfrac{{\vec {\alpha} \vec
{p}_{f}} }{{E\left( {\vec {p}_{f}}  \right) + m}}\alpha ^{k}\! +
\alpha ^{k}\!\dfrac{{\vec {\alpha} \left( {\vec {p}_{i} + \vec
{k}} \right)}}{{E\left( {\vec {p}_{i} + \vec {k}} \right) + m}}}
\right)\!,\quad \mu = k,\; k = 1 ,2, 3}; \\
 \end{array} \right.\\
 &&\!\!\!\!\!\!\!\!\!\!N^{\mu} \left( {\vec {p}_{f} ; \vec {p}_{i} + \vec
{k}} \right) = \left\{ \begin{array}{l} { e  R_{f}  R_{1} \beta
\left( {\dfrac{{\vec {\alpha} \vec {p}_{f}} }{{E\left( {\vec
{p}_{f}}  \right) + m}}} - \dfrac{{\vec {\alpha }\left( {\vec
{p}_{i} + \vec {k}} \right)}}{{E\left( {\vec {p}_{i} + \vec {k}}
\right) + m}}\right),\quad \mu = 0;} \\ {  - R_{f}  R_{1} \left(
{\alpha ^{k} - \dfrac{{\vec {\alpha} \vec {p}_{f}} }{{E\left(
{\vec {p}_{f}}  \right) + m}}\alpha ^{k}\dfrac{{\vec {\alpha}
\left( {\vec {p}_{i} + \vec {k}} \right)}}{{E\left( {\vec {p}_{i}
+ \vec {k}} \right) + m}}} \right),\quad \mu = k,\; k = 1,2,3;}
 \end{array} \right.  \end{eqnarray*}
 $R_{f} \equiv R_{p_{f}}$, $R_{1} \equiv R_{p_{i} + k} $
etc. Functions (\ref{eq4}) with relations (\ref{eq5}) were used to
obtain $S_{fi}$  in the impulse representation.

If a special gauge is taken, in which the initial and final
photons are transversely polarized in the laboratory frame $\left(
{\vec {p}_{i} = 0;\quad \varepsilon ^{0} = \varepsilon ^{'0} =
0;\quad \vec {\varepsilon }\vec {k} = \vec {\varepsilon} '\vec
{k}' = 0} \right)$, then the expression for $S_{fi}$
 is simplified:
\[ A = \dfrac{{e^{2}}}{{m}}  R_{f} \left( \vec {\varepsilon}
 {\vec {\varepsilon} }^{\,'} + \dfrac{{\vec {\sigma} \left(
{\vec {k} - {\vec {k}}^{'}} \right)  \vec {\sigma} {\vec
{\varepsilon} }^{\,'}  \vec {\sigma} \vec {k}  \vec {\sigma} \vec
{\varepsilon}}}{2k_{0}\left( {2m + k_{0} - k_{0}^{\prime} }
\right)} + \dfrac{{\vec {\sigma} \left( {\vec {k} - {\vec
{k}}^{'}} \right)  \vec {\sigma} \vec {\varepsilon}  \vec {\sigma
}{\vec {k}}^{'}  \vec {\sigma} {\vec {\varepsilon}
}^{\,'}}}{2k_{0}^{\prime }\left( {2m + k_{0} - k_{0}^{\prime} }
\right)}\right).\]

Then Klein-Nishina-Tamm formula for Compton effect differential
cross section can be obtained with routine methods.

\subsection*{4\quad Electron self-energy}
\begin{eqnarray*}
&& - i\sum {^{\left( {2} \right)}}\left( {p} \right) = - \int
\dfrac{{d^{4}k}}{{\left( {2\pi}  \right)^{4}  k^{2}}}\times\\
&&\times \left[ {K_{1}^{\mu }} \left( {\vec {p}; \vec {p} - \vec
{k}; \nu = - 1} \right)\dfrac{{1}}{{p_{0} - k_{0} - \beta E\left(
{\vec {p} - \vec {k}} \right)}}(K_{1})_\mu  \left( {\vec {p} -
\vec {k}; \vec {p};{\nu}^{'} = 1} \right) - \right.\\
 &&- K_{1}^{\mu}  \left( {\vec {p}; \vec {p} - \vec {k}; \nu = - 1}
\right)\dfrac{{\dfrac{{1}}{{2}}\left( {I+\beta} \right)}}{{\beta
E\left( {\vec {p}} \right) - k_{0} - E\left( {\vec {p} - \vec {k}}
\right)}}  (K_{1})_\mu   \left( {\vec {p} - \vec {k}; \vec {p};
{\nu}^{'} = 1} \right) +
\\
&& + C^{\mu} \left( {\vec {p}; \vec {p} - \vec {k}}
\right)\dfrac{{\dfrac{{1}}{{2}}\left( {I+\beta} \right)}}{{\beta
E\left( {\vec {p}} \right) - k_{0} - E\left( {\vec {p} - \vec {k}}
\right)}}  C_{\mu}  \left( {\vec {p} - \vec {k};\vec {p}} \right)
+ \\
 && \left.+ N^{\mu} \left( {\vec {p}; \vec {p} - \vec {k}}
\right)\dfrac{{\dfrac{{1}}{{2}}\left( {I-\beta}  \right)}}{{ -
\beta E\left( {\vec {p}} \right) - k_{0} + E\left( {\vec {p} -
\vec {k}} \right)}}N_{\mu}  \left( {\vec {p} - \vec {k}; \vec {p}}
\right) \right].
 \end{eqnarray*}

For $p^{2} = m^{2}$, with taking into account that for electrons
$\beta  \psi _{FW}^{\left( { +}  \right)} =
 \psi _{FW}^{\left( { + } \right)} $, we obtain $ - i\dsum {^{\left(
{2} \right)}\left( {p} \right)  = } - \dfrac{{2e^{2}}}{{E\left(
{\vec {p}} \right)}}\int {\dfrac{{d^{4}k}}{{\left( {2\pi}
\right)^{4}  k^{2}}}}  \dfrac{{p  k + m^{2}}}{{\left[ {\left( {p -
k} \right)^{2} - m^{2}} \right]}}$, which is the same as the
expression for the mass operator in the Dirac representation with
taking into consideration the normalization of spinors in the
external electron lines.

\subsection*{5\quad Vacuum polarization}

The diagram in Fig.~5 is correspondent with the following
expression for the polarization operator:
\begin{eqnarray*}
&& \Pi^{\mu \nu} \left( {q} \right) = i\int {\dfrac{{d\vec
{p}}}{{\left( {2\pi} \right)^{3}}}} \left\{
{Sp\dfrac{{1}}{{E\left( {\vec {p}} \right) - q^{0} + E\left( {\vec
{p} - \vec {q}} \right)}}} N^{\mu} \left( {\vec {p} - \vec {q};
\vec {p}} \right)\dfrac{{I+\beta} }{{2}}N^{\nu} \left( {\vec {p};
\vec {p} - \vec {q}} \right) +\right. \\ && + \left.
Sp\dfrac{{1}}{{E\left( {\vec {p}} \right) + q^{0} + E\left( {\vec
{p} - \vec {q}} \right)}}N^{\mu} \left( {\vec {p} - \vec {q}; \vec
{p}} \right)\dfrac{{I-\beta} }{{2}}N^{\nu} \left( {\vec {p}; \vec
{p} - \vec {q}} \right) \right\}
 \end{eqnarray*}

On the spur calculation the expression for $\left( { - i}
\right)\Pi^{\mu \nu }$ is the same as Heitler induction tensor
$L^{\mu \nu} $ \cite{5}.

\subsection*{6\quad Radiation corrections to electron scattering\\
in the external field}

When calculating radiation corrections from the diagrams of
Fig.~6, it turns out that the matrix elements $S_{fi} $
corresponding to the electron-positron propagator diagrams cancel
out with the matrix element parts corresponding to the
propagatorless diagrams $d)$, $h)$, $l)$ in Fig.~6. As a result,
with account for the limiting Heitler process for singular
denominators \cite{5}, the matrix element for the desired
radiation corrections is
\begin{eqnarray*}
&& S_{fi} = \dfrac{{1}}{{\left( {2\pi}  \right)^{3}}}\left( {\chi
_{s_{f}} ^{ + }\quad 0} \right)\left\{ \int\dfrac{d^4k}{(2\pi)^4
k^2} \int {\dfrac{d\varepsilon \delta \left( {\varepsilon}
\right)}{\vec {p}_{f}^{2} \varepsilon \left( {2 + \varepsilon}
\right)}}\right. \Bigg[ C_\mu \left( \vec p_f;\vec p_f\left(
1+\varepsilon \right) -\vec k\right) \times  \\
&&\times\dfrac{E\left( \vec p_f\right) +E(\vec p_f(1+\varepsilon
))}{E\left( \vec p_f\right) -k_0-E(\vec p_f\left( 1+\varepsilon
\right) -\vec k)}C^\mu (\vec p_f\left( 1+\varepsilon \right) -\vec
k;\vec p_f\left( 1+\varepsilon \right) )+ N_\mu (\vec p_f;\vec
p_f\left( 1+\varepsilon \right) -\vec k)\times \\ &&\times \left.
\dfrac{E\left( \vec p_f\right) +E(\vec p_f(1+\varepsilon
))}{E\left( \vec p_f\right) -k_0+E(\vec p_f\left( 1+\varepsilon
\right) -\vec k)}N^\mu (\vec p_f\left( 1+\varepsilon \right) -\vec
k;\vec p_f\left( 1+\varepsilon \right) )\right]   C^\nu \left(
\vec p_f;\vec p_i\right) A_\nu \left( q\right) + \\ &&+\int
\dfrac{d^4k}{\left( 2\pi \right) ^4 k^2}\int \dfrac{d\varepsilon
\delta \left( \varepsilon \right) }{\vec p_i^{\,2}\varepsilon
\left( 2+\varepsilon \right) }\Bigg[ C_\mu \left( \vec p_i\left(
1+\varepsilon \right) ;\vec p_i\left( 1+\varepsilon \right) -\vec
k\right) \times  \\ &&\times\dfrac{E\left( \vec p_i\right) +E(\vec
p_i(1+\varepsilon ))}{E\left( \vec p_i\right) -k_0-E(\vec
p_i\left( 1+\varepsilon \right) -\vec k)} C^\mu (\vec p_i\left(
1+\varepsilon \right) -\vec k;\vec p_i)+ N_\mu (\vec
p_i(1+\varepsilon );\vec p_i\left( 1+\varepsilon \right) -\vec
k)\times\\ &&\times \left. \dfrac{E\left( \vec p_i\right) +E(\vec
p_i(1+\varepsilon ))}{E\left( \vec p_i\right) -k_0+E(\vec
p_i\left( 1+\varepsilon \right) -\vec k)} N^\mu (\vec p_i\left(
1+\varepsilon \right) -\vec k;\vec p_i)\right]   C^\nu \left( \vec
p_f;\vec p_i\right) A_\nu \left( q\right) - \\ &&-\int
\dfrac{d^4k}{\left( 2\pi \right) ^4 k^2}\dfrac 1{2E\left( \vec
p_f\right) }\left[ \dfrac 1{E\left( \vec p_f\right) -k_0+E(\vec
p_f-\vec k)}N_\mu \left( \vec p_f;\vec p_f-\vec k\right) C^\mu
\left( \vec p_f-\vec k;\vec p_f\right) +\right.  \\ &&+\left.
\dfrac 1{E\left( \vec p_f\right) -k_0-E(\vec p_f-\vec k)}C_\mu
\left( \vec p_f;\vec p_f-\vec k\right) N^\mu \left( \vec p_f-\vec
k;\vec p_f\right) \right]  N^\nu \left( \vec p_f;\vec p_i\right)
A_\nu \left( q\right) - \\&& -\int \dfrac{d^4k}{\left( 2\pi
\right) ^4 k^2}\dfrac 1{2E\left( \vec p_i\right) }N^\nu \left(
\vec p_f;\vec p_i\right) A_\nu \left( q\right) \left[ \dfrac
1{E\left( \vec p_i\right) -k_0-E(\vec p_i-\vec k)}\right. \times\\
 &&\times N_{\mu}  \left( {\vec {p}_{i} ;\vec {p}_{i} - \vec {k}}
\right) C^{\mu} \left( {\vec {p}_{i} - \vec {k}_{i} ;\vec {p}_{i}}
\right) + \dfrac{{1}}{{E\left( {\vec {p}_{i}}  \right) - k_{0} +
E\left( {\vec {p}_{i} - \vec {k}} \right)}} \times \\
 &&\times C_{\mu}  \left( {\vec {p}_{i} ; \vec {p}_{i} - \vec {k}}
\right)N^{\mu} \left( {\vec {p}_{i} - \vec {k}_{i} ; \vec
{p}_{_{i}} } \right) \Bigg] - \\
 &&- \int {\dfrac{{d^{4}k}}{{\left( {2\pi}  \right)^{4}  k^{2}}}} \left[
{C^{\mu} \left( {\vec {p}_{f} ; \vec {p}_{f} - \vec {k}}
\right)\dfrac{{1}}{{E\left( {\vec {p}_{f}}  \right) - k_{0} -
E\left( {\vec {p}_{f} - \vec {k}} \right)}}} \right.C^{\nu} \left(
{\vec {p}_{f} - \vec {k}; \vec {p}_{i} - \vec {k}} \right)A_{\nu}
\left( {q} \right) \times \\
 &&\times \dfrac{{1}}{{E\left( {\vec {p}_{i}}  \right) - k_{0} - E\left( {\vec
{p}_{i} - \vec {k}} \right)}}C_{\mu}  \left( {\vec {p}_{i} - \vec {k};\vec
{p}_{i}}  \right) + \\
 &&+ N^{\nu} \left( {\vec {p}_{f} ;\vec {p}_{f} - \vec {k}}
\right)\dfrac{{1}}{{E\left( {\vec {p}_{f}}  \right) - k_{0} +
E\left( {\vec {p}_{f} - \vec {k}} \right)}}  C^{\nu}\left( {\vec
{p}_{f} - \vec {k}; \vec {p}_{i} - \vec {k}} \right)A_{\nu} \left(
{q} \right) \times \\
 &&\times \dfrac{{1}}{{E\left( {\vec {p}_{i}}  \right) - k_{0} + E\left( {\vec
{p}_{i} - \vec {k}} \right)}}N_{\mu}  \left( {\vec {p}_{i} - \vec {k};\vec
{p}_{i}}  \right) + \\
 &&+  N^{\mu} \left( {\vec {p}_{f} ;\vec {p}_{f} - \vec {k}}
\right)\dfrac{{1}}{{E\left( {\vec {p}_{f}}  \right) - k_{0} +
E\left( {\vec {p}_{f} - \vec {k}} \right)}}  N^{\nu}\left( {\vec
{p}_{f} - \vec {k}; \vec {p}_{i} - \vec {k}} \right)A_{\nu} \left(
{q} \right) \times \\
 &&\times \dfrac{{1}}{{E\left( {\vec {p}_{i}}  \right) - k_{0} - E\left( {\vec
{p}_{i} - \vec {k}} \right)}}C_{\mu}  \left( {\vec {p}_{i} - \vec {k};\vec
{p}_{i}}  \right) + \\
 &&+ C^{\mu} \left( {\vec {p}_{f} ; \vec {p}_{i} - \vec {k}}
\right)\dfrac{{1}}{{E\left( {\vec {p}_{f}}  \right) - k_{0} -
E\left( {\vec {p}_{f} - \vec {k}} \right)}}N^{\nu}\left( {\vec
{p}_{f} - \vec {k}; \vec {p}_{i} - \vec {k}} \right)A_{\nu} \left(
{q} \right) \times \\
 &&\times \left. {\dfrac{{1}}{{E\left( {\vec {p}_{i}}  \right) - k_{0} +
E\left( {\vec {p}_{i} - \vec {k}} \right)}}  N_{\mu}  \left( {\vec
{p}_{i} - \vec {k}; \vec {p}_{i}}  \right)} \right] - \\ && -
C_{\mu}  \left( {\vec {p}_{f} ;\vec {p}_{i}}
\right)\left.\dfrac{{\Pi^{\mu \nu }\left( {q} \right)}}{{q^{2}}}
A_{\nu} \left( {q} \right)\right\}  \left( \begin{array}{c}
 {\chi _{s_{i}} } \\
 {0}
\end{array}\right).
 \end{eqnarray*}

Upon the electron mass and charge renormalization, the written
matrix element $S_{fi} $ can be used to calculate anomalous
magnetic moment of the electron and Lamb shift of energy atomic
levels. The final results of the calculations agree with those in
the Dirac representation.
\subsection*{7\quad Electron-positron pair annihilation}

In the second order of the perturbation theory the
electron-positron pair annihilation is correspondent with the
diagrams of Fig.~3 with substitution $\varepsilon ,k \to
\varepsilon _{1}, - k_{1}$;  ${\varepsilon} ',{k}' \to \varepsilon
_{2} ,k_{2}$;  $p_{i} ,s_{i} \to p_{ -} ,s_{ -}$; $p_{f} ,s_{f}
\to - p_{ + } ,s_{ +}  $.

With account for the extension to eight dimensions, matrix element
$S_{ + - } $ of the process is  $$ S_{ + -}  = \dfrac{{ - i\left(
{2\pi} \right)^{4}\delta ^{4}\left( {p_{ -} + p_{ +}  - k_{1} -
k_{2}} \right)}}{{\left( {2\pi} \right)^{3}  \sqrt {2k_{1}^{0}
 2k_{2}^{0}} } }\left(  {0\quad V_{s_{ +} }^{ +} }
\right)\dfrac{{1}}{{2}}A_{1} \left(
\begin{array}{c}
 {U_{s_{ -} } }  \\
 {0}
 \end{array}\right).
$$

With account for the above substitution, operator $A_{1} $ is the
same in its structure as operator $A$ in the expression of $S_{fi}
$ for Compton effect, with $C^{\mu}  \to C_{1}^{\mu}  , N^{\mu}
\to N_{1}^{\mu}  ,$
\begin{eqnarray*}
&&C_{1}^{\mu}  ( { - \vec {p}_{ +}  ; \vec {p}_{ -}  - \vec
{k}_{1}} ) = \left\{ \begin{array}{l}
 {e  R_{p_{ +} }}
 \left(I+\rho-\dfrac{\beta\vec{\alpha}(-\vec{p}_{+})}
 {E\left( { - \vec {p}_{ +} }  \right) + \beta _{1}  m}
 {\left( {I + \rho}  \right)}\right.\times\\
 \qq\qq\times\left.\dfrac{{\beta \vec {\alpha} \left(
{\vec {p}_{ -}  - \vec {k}_{1}}  \right)}}{{E\left( {\vec {p}_{ -}
- \vec {k}_{1}}  \right) + \beta _{1}  m}} \right)R_{p_{-}-k_1},\q
\mu =0;
\\
-eR_{p_{+}}\left( \dfrac{\beta \vec \alpha \left( -\vec p_{+}\right) }{%
E\left( -\vec p_{+}\right) +\beta _1 m}\alpha ^k\left( I+\rho
\right) -\left( I+\rho \right) \alpha ^k\times\right.\\
\qq\qq\times\left.\dfrac{\beta \vec \alpha \left( \vec
p_{-}-\vec k_1\right) }{E\left( \vec p_{-}-\vec k_1\right) +\beta _1 m}%
\right) R_{p_{-}-k_1},\q\mu =k,\;k=1,2,3;
 \end{array}\right.\\
&&N_1^\mu (-\vec p_{+};\vec p_{-}-\vec k_1)=\left\{
\begin{array}{l}
eR_{p_{+}}\left( \dfrac{\beta \vec \alpha \left( -\vec
p_{+}\right) }{E\left( -\vec p_{+}\right) +\beta _1 m}\left(
I+\rho \right) -\left( I+\rho \right) \times\right.\\
\qq\qq\times\left.\dfrac{\beta \vec \alpha \left( \vec p_{-}-\vec
k_1\right) }{E\left( \vec p_{-}-\vec k_1\right) +\beta _1
m}\right) R_{p_{-}-k_1},\q\mu =0; \\
 -eR_{p_{+}}\left( \alpha
^k\left( I+\rho \right) -\dfrac{\beta \vec \alpha
\left( -\vec p_{+}\right) }{E\left( -\vec p_{+}\right) +\beta _1 m}%
\left( I+\rho \right) \alpha ^k\times\right.\\
\qq\qq\times\left.\dfrac{\beta \vec \alpha \left( \vec p_{-}-\vec
k_1\right) }{E\left( \vec p_{-}-\vec k_1\right) +\beta _1
m}\right) R_{p_{-}-k_1},\q\mu =k,\;k=1,2,3.
\end{array}
\right.
 \end{eqnarray*}

The expression for $S_{ + -}  $ allows us to obtain the
differential cross section of the electron-positron pair
annihilation, which is the same as that in the Dirac
representation.

\end{document}